\documentclass[10pt,journal,compsoc]{IEEEtran}

\def\X{{\cal X}}

\def\C{{\cal C}}

\def\T{{\cal T}}
\def\N{{\cal N}}

\usepackage{amsthm}
\theoremstyle{plain}
\newtheorem{theorem}{Theorem}[section]
\newtheorem*{corollary}{Corollary}

\theoremstyle{definition}
\newtheorem*{definition}{Definition}

\usepackage{amsfonts}
\usepackage{amssymb}

\usepackage{mathtools}
\DeclarePairedDelimiter{\ceil}{\lceil}{\rceil}

\usepackage{nccmath}

\newcounter{algorithm}
\newcommand{\algcaption}[1]{%
    \par\noindent
    \refstepcounter{algorithm}%
    \textbf{Algorithm \thealgorithm:}
    #1\par
}

\usepackage{multirow}
\usepackage[flushleft]{threeparttable}
\ifCLASSOPTIONcompsoc
  \usepackage[nocompress]{cite}
\else
  \usepackage{cite}
\fi

\ifCLASSINFOpdf
  \usepackage[pdftex]{graphicx}
  \graphicspath{{figs/}}
\else
  \usepackage[dvips]{graphicx}
  \graphicspath{{../eps/}}
\fi

\usepackage{amsmath}
\usepackage{algpseudocode}

\ifCLASSOPTIONcompsoc
  \usepackage[caption=false,font=footnotesize,labelfont=sf,textfont=sf]{subfig}
\else
  \usepackage[caption=false,font=footnotesize]{subfig}
\fi
\usepackage{url}

\begin{document}

\title{Distribution Agnostic Symbolic Representations for Time Series Dimensionality Reduction and Online Anomaly Detection}
%
%
%
%

\author{Konstantinos~Bountrogiannis,~\IEEEmembership{}
        George~Tzagkarakis,~\IEEEmembership{}
        Panagiotis~Tsakalides,~\IEEEmembership{Member,~IEEE}
\thanks{This work is accepted for publication in the IEEE Transactions on Knowledge and Data Engineering. This is the final preprint version. For the published version, refer to https://ieeexplore.ieee.org/document/9774017}

}

\IEEEtitleabstractindextext{%
\begin{abstract}
Due to the importance of the lower bounding distances and the attractiveness of symbolic representations, the family of symbolic aggregate approximations (SAX) has been used extensively for encoding time series data. However, typical SAX-based methods rely on two restrictive assumptions; the Gaussian distribution and equiprobable symbols. This paper proposes two novel data-driven SAX-based symbolic representations, distinguished by their discretization steps. The first representation, oriented for general data compaction and indexing scenarios, is based on the combination of kernel density estimation and Lloyd-Max quantization to minimize the information loss and mean squared error in the discretization step. The second method, oriented for high-level mining tasks, employs the Mean-Shift clustering method and is shown to enhance anomaly detection in the lower-dimensional space. Besides, we verify on a theoretical basis a previously observed phenomenon of the intrinsic process that results in a lower than the expected variance of the intermediate piecewise aggregate approximation. This phenomenon causes an additional information loss but can be avoided with a simple modification. The proposed representations possess all the attractive properties of the conventional SAX method. Furthermore, experimental evaluation on real-world datasets demonstrates their superiority compared to the traditional SAX and an alternative data-driven SAX variant.
\end{abstract}

}

\IEEEoverridecommandlockouts
\IEEEpubid{\begin{minipage}{\textwidth}\copyright\ 2022 IEEE. Personal use of this material is permitted. Permission from IEEE must be obtained for all other uses, in any current or future media, including reprinting/republishing this material for advertising or promotional purposes, creating new collective works, for resale or redistribution to servers or lists, or reuse of any copyrighted component of this work in other works. \hfill \hspace{\columnsep}\end{minipage}}

\maketitle
\IEEEpubidadjcol

\IEEEdisplaynontitleabstractindextext

%
\IEEEpeerreviewmaketitle

\IEEEraisesectionheading{\section{Introduction}\label{sec:introduction}}

%
%
%

\IEEEPARstart{D}{ata} representation in a lower-dimensional space is a central topic in storing and mining the ever increasing amount of time series becoming available thanks to the advances in computing and sensing technologies. Certainly, the extraction of descriptive motifs of reduced dimensionality, which provide a meaningful, yet compact, representation of the original inherent information, has been an ongoing research field for the last few decades.

At the core of time series representations is the definition of appropriate distance measures in the lower-dimensional space. More precisely, the transformation of data objects to a lower-dimensional subspace should be, ideally, distance-preserving. This property would allow data mining tasks to perform equally well in the lower-dimensional space as in the higher-dimensional one. However, except for trivial cases, distance measures defined in lower-dimensional spaces can only approximately preserve distances.

A milestone for the definition and utilization of such measures has been the discovery of the properties of lower-bounding distances~\cite{bib:GEMINI}. That is, when the distance measure in the lower-dimensional space lower-bounds the distance of the same objects in the higher-dimensional space, it is proved that approximate queries in the lower-dimensional space (i.e., queries that return all objects with up to a maximum distance from the query object), return no false dismissals. In advance, the tighter this bound is, the less false alarms are returned.

The symbolic aggregate approximation (SAX) introduced in~\cite{bib:SAX} is the first symbolic representation to possess the lower-bounding property. Symbolic representations have the advantage of exploiting the wealth of search algorithms used in bioinformatics and text mining communities, in conjunction to enabling the exploitation of numerous data mining techniques that require discrete data. On top of that, SAX is characterized by conceptual simplicity and computational tractability. These properties have rendered SAX arguably the most widely used time series representation for monitoring, processing and mining data from numerous sources and in distinct application domains, including physiological data~\cite{bib:Ord11}, space telemetry~\cite{bib:Chiu03}, computational biology~\cite{bib:Andr05}, smart grids~\cite{bib:Wang16}, building systems~\cite{bib:Mil15}, and stock market~\cite{bib:Agha14}, just to name a few.

Several variants have been proposed to enhance the performance of conventional SAX. Specifically, the symbolic representations proposed in~\cite{bib:TSX, bib:1d-SAX} incorporate trend information but they lack the lower-bounding property. This limitation is overcome in~\cite{bib:TFSAX, bib:SAX-TD}, where the corresponding symbolic representations exploit trend information whilst retaining the lower-bounding property. Nevertheless, trend information increases the dimensionality of the representations, thus it is not clear whether the quality of the distance measures is indeed higher or not when the target dimensionality is fixed. Furthermore, the overall compressibility deteriorates, as the extra dimensions are mostly continuous, not discretized. The same holds for the representations introduced by~\cite{bib:SAX_SD} and~\cite{bib:SAX_Stat}, which incorporate statistical features in the representation. The distribution of the time series is considered in~\cite{bib:asax} and~\cite{bib:psax} for optimizing the discretization step without increasing the output dimensionality. Lastly, the paper~\cite{bib:SAX_lookup} exploits the minimum distance between adjacent symbols to improve the distance measure of SAX representations.

This paper focuses on improving the discretization step of SAX. First, it extends the work in~\cite{bib:psax} that introduced a probabilistic SAX (pSAX) variant -- tailored to general data compaction and indexing scenarios -- based on a kernel density estimator (KDE)~\cite{bib:KDE} to estimate the density function of the data source, coupled with a Lloyd-Max quantizer~\cite{bib:Lloyd} for computing optimal discretization intervals. In particular, we now concern i) a thorough information-theoretic analysis of SAX, in terms of information loss, to quantify the information retained by pSAX compared to SAX, and ii) a more extensive experimental evaluation of pSAX. Second, this paper proposes a novel symbolic representation, hereafter referred to as clustering SAX (cSAX), that relies on the mean-shift clustering method~\cite{bib:Mean-Shift02} to produce descriptive symbolic sequences that are more appropriate for high-level data analysis tasks. Notably, our pSAX and cSAX representations can be combined with the techniques in the other SAX-based representations, as far as they do not alter the discretization step (ref. 
\cite{bib:TSX} --~\cite{bib:UncertaintySAX}).

The contribution of this paper also comprises: i) a theoretical verification, along with a practical solution, of a previously observed phenomenon, where the data variance after the intermediate step of piecewise aggregate approximation is lower than the expected~\cite{bib:SAXequiprob}; ii) a dynamic clustering criterion for the mean-shift method; and iii) the definition of two novel distance measures, which are not lower-bounding, but are optimal in the sense of mean squared error. The latter is of importance in data mining applications, where there is no need to identify the time series within a strict distance of the query. In those cases, the lower-bounding property is irrelevant and a minimum-error distance measure is more effective~\cite{bib:Wang08}.

The proposed pSAX and cSAX methods are non-parametric and data-driven, and are shown to effectively exploit the underlying distribution of the data towards increasing the accuracy of the representations. Both methods rely on kernel-based techniques to estimate the underlying data distribution. Importantly, both methods preserve the lower-bounding property, whilst pSAX exhibits tighter lower bounds than the conventional SAX. Furthermore, both methods are experimentally compared against the conventional SAX and the aSAX variant~\cite{bib:asax}, which employs the \mbox{$k$-means} algorithm for the discretization step.

The rest of the paper is organized as follows: Section~\ref{sec:prelim} overviews the SAX technique, along with the anomaly detection methods utilized for testing cSAX. Section~\ref{sec:information} quantifies the information loss of SAX and provides guidelines for its minimization. Section~\ref{sec:represent_prelim} analyzes the key components of our proposed symbolic representations, whilst Sections~\ref{sec:psax} and~\ref{sec:csax} provide the implementation details of pSAX and cSAX, respectively. Section~\ref{sec:results} evaluates the performance of our proposed methods on distinct time series data. Finally, Section~\ref{sec:concl} summarizes the main outcomes and gives directions for further work.


\section{Preliminaries}\label{sec:prelim}

\subsection{Symbolic Aggregate Approximation}\label{sec:sax}

This section overviews briefly the main concepts of the conventional SAX method~\cite{bib:SAX}. The core of SAX consists of a two-step transformation that reduces the dimensionality of the data, coupled with the definition of an appropriate distance measure in the lower-dimensional space that lower bounds the Euclidean distance in the original space.

Let $\T^N = \{X:X=(x_1,x_2,\dots,x_N)\}$ denote the set of all $N$-length continuous (real-valued) time series objects and $\C^N_A = \{C:C=(c_1,c_2,\dots,c_N),c_i \in A\}$ the set of all $N$-length discrete time series objects with values from an alphabet $A$ of size $|A|=\kappa$. Given a time series $X\in \T^N$, a Z-normalization is performed as a preprocessing step prior to the application of a SAX transformation,
\begin{equation}
    X' = \frac{X-\mu_X}{\sigma_X}\ ,
\end{equation}
where $\mu_X$ and $\sigma_X$ are the mean and standard deviation of $X$, respectively. For convenience, hereafter $X$ is assumed to be Z-normalized {\`a} priori.

\begin{figure}
    \centering
    \includegraphics[width=.99\linewidth]{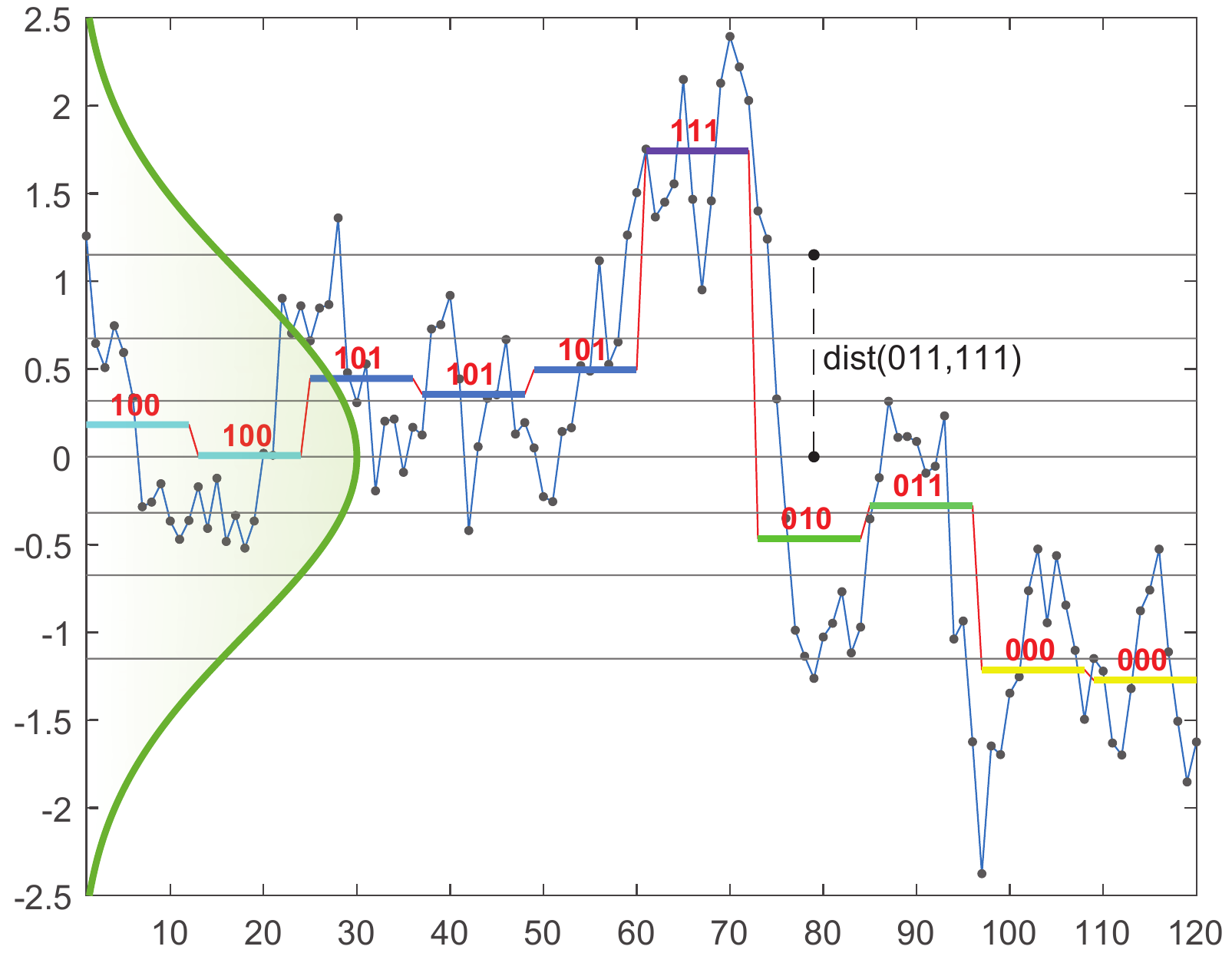}
    \caption{SAX representation of a time series. In this example, a time series of length $N=120$ is first transformed into its PAA representation by segmenting and averaging the series into $M=12$ pieces. Then, each segment is assigned a binary codeword, subject to which of the $\kappa=8$ equiprobable intervals of the standard Gaussian pdf it falls in. Each quantization interval is bounded by two cutlines and is assigned a codeword from the alphabet $A=\{0_2,1_2,\dots,7_2\}$.}
    \label{fig:SAX}
\end{figure}

The first step of SAX consists of a piecewise aggregate approximation (PAA) $\T^N \rightarrow \T^M$, which transforms a given time series $X\in \T^N$ into another time series $Y=(y_1,\dots,y_M)\in\T^M$ with reduced dimensionality $M < N$. For this, $X$ is divided into $M$ equal-sized segments and the average value is calculated for each segment. The second step performs a discretization $\T^M \rightarrow \C^M_A$ that maps $y_i\in\mathbb{R}$ to $c_i\in A$. The discretization scheme adopted by SAX employs an equiprobable quantization of $y_i$, $i=1,\dots,M$, under the assumption that $y_i\sim \N(0,1)$. Under this assumption, $y_i$'s are quantized within $\kappa$ equiprobable intervals under the Gaussian probability density curve. The two-step transformation $\T^N \rightarrow \T^M \rightarrow \C^M_A$ results in a SAX representation of length $M$ for the alphabet $A$. The ratio $M/N$ determines the degree of dimensionality reduction. Fig.~\ref{fig:SAX} illustrates the two-step process of SAX.

It is known that, in order to guarantee the absence of false dismissals when performing a similarity search in the lower-dimensional space, it suffices to define a distance measure in the lower-dimensional space that lower bounds the distance in the original space (ref.~\cite{bib:GEMINI}). Let $C,\,Q$~$\in\C^M_A$ be the SAX representations of two time series $U\in\T^N$ and $S\in\T^N$, respectively. Then, a lower-bounding distance measure is defined as follows,
\begin{equation}\label{eq:mindist}
    mindist(C,Q) = \sqrt{\frac{N}{M}\sum_{i=1}^M \left(dist(c_i,q_i)\right)^2}\ ,
\end{equation}
where $dist(c_i,q_i)$ is the absolute difference of the two closest cutlines that respectively bound the intervals of $c_i$ and $q_i$ (ref. Fig.~\ref{fig:SAX} for an example). Furthermore, if $Y\in\T^M$ is the PAA of $U$ and \mbox{$Q\in\C^M_A$} is the SAX representation of $S$, a tighter lower bounding distance measure is defined by
\begin{equation}\label{eq:mindist_PAA}
    mindist\_PAA(Y,Q)=\hspace{-3.9pt}
            \sqrt{\frac{N}{M}\sum_{i=1}^M
            \begin{cases}
            (\beta_{L_i}-y_i)^2,\enskip \beta_{L_i}>y_i\\
            (\beta_{U_i}-y_i)^2,\enskip \beta_{U_i}<y_i\\
            \text{\qquad} 0, \text{\qquad otherwise}
            \end{cases}}
\end{equation}
where $\beta_{L_i}$ and $\beta_{U_i}$ are the lower and upper cutlines of the codeword $q_i$.

The tightness of lower bound (TLB) is the degree of closeness of the $mindist\_PAA(Y,Q)$ to the true distance of the data. The higher it is, the lower the rate of false positives in a similarity search. The TLB is defined as follows,
\begin{equation}\label{eq:TLB}
    TLB(U,S) = \frac{mindist\_PAA(Y,Q)}{d(U,S)}\ ,
\end{equation}
where $d(U,S)$ is the distance in the original time series space.

Because the discretization step $\T^M \rightarrow \C^M_A$ quantizes the PAA samples into $\kappa$ equiprobable intervals, when the Gaussian assumption is valid, the generated symbolic sequence is uniformly distributed.

Although not stated clearly in the introductory paper~\cite{bib:SAX}, the selection of equiprobable intervals is typically based on the fact that the uniform distribution maximizes the entropy of the symbolic sequence. This implies that a pair of time series is more easily distinguishable in the lower-dimensional space.

An advantage of our proposed method is that the discretization step is optimized for general data mining applications, such that the mean squared error (MSE) is minimized. Moreover, we demonstrate that MSE minimization yields an improved performance when compared against entropy maximization, in terms of TLB~\eqref{eq:TLB}.

\subsection{Anomaly Detection via Statistical Hypothesis Testing}\label{sec:anomaly}

A non-parametric method for online anomaly detection has been proposed in~\cite{bib:KL_anomaly}. The method proceeds in a rolling window fashion, employing a goodness-of-fit criterion to test whether the most recent window of $n$ samples is distributed similarly with the past samples. In particular, the null hypothesis of the test is a composite hypothesis consisting of the union of empirical distributions for a collection of past $n$-length windows. Then, the null hypothesis is partitioned into multiple simple hypotheses, which are tested separately. In practice, a similarity test is performed between the current window and each previous window independently. A similarity condition is assumed to be satisfied if the current window fits with the distribution of at least one of the previous windows. If the null hypothesis is rejected, the current window is flagged as anomalous, whilst if the null hypothesis is accepted, the current window is flagged as normal. The above process is summarized in Algorithm~\ref{alg:anomaly}.

An interesting observation regarding Algorithm~\ref{alg:anomaly} is that the earliest windows will always be considered as anomalous, as there will be very few pieces in the null hypothesis to test against. At this stage, the method learns the most diverse representations of the data and saves them as part of the null hypothesis. Gradually, the null hypothesis makes a good representation of the data and acts as a ground truth model to test the goodness of fit for future windows.

~
\algcaption{Anomaly Detection via Goodness of Fit\\
\textbf{Input:} $x$ (time series), $n$ (window length), $\alpha$ (significance level)}\label{alg:anomaly}
\begin{algorithmic}[1]
\State $\rho_{thr} = F_{T,n}^{-1}(1-\alpha)$
\State $c \gets 0$ \Comment{null hypothesis count}
\State $i \gets n$ \Comment{sample index}
\While{new data arrives}
    \If{$T(x[i\! -\! n \! +\! 1\! : \! i],H_{0j}) < \rho_{thr}$, for any null hypothesis component $H_{0j},\ j\leq c$}
        \State Declare window as non-anomalous
    \Else
        \State Declare window as anomalous
        \State $c \gets c+1$
        \State $H_{0c} \gets x[i\! -\! n \! +\! 1\! : \! i]$
    \EndIf
    \State $i \gets i+1$
\EndWhile
\end{algorithmic}
~

Furthermore, as stated in~\cite{bib:KL_anomaly}, the specific test statistic, $T$, employed by Algorithm~\ref{alg:anomaly} is defined by:
\begin{equation}\label{eq:statistic}
    T(X,Q) \triangleq 2\cdot n\cdot D_{KL}(P_X\!\parallel\! P_Q)\ ,
\end{equation}
where $P_X$ and $P_Q$ are the empirical mass functions of $X$ and $Q$, respectively, and $D_{KL}(P_X\!\parallel\! P_Q)$ is the Kullback-Leibler divergence between the mass functions $P_X$ and $P_Q$:
\begin{equation}\label{eq:KL}
    D_{KL}(P_X\!\parallel\! P_Q) \triangleq \sum_{x\in\X} P_X(x)\log{\frac{P_X(x)}{P_Q(x)}}\ ,
\end{equation}
with the convention that $0\log{0}=0$.

It is known that $T(X,Q)$ converges in distribution to a random variable that follows the $\chi^2$ distribution with $\kappa-1$ degrees of freedom, where $\kappa$ is the cardinality of the sample space. In light of this observation, the threshold $\rho_{thr}$ in Algorithm~\ref{alg:anomaly} is computed by the inverse cumulative distribution function of the $\chi^2$ distribution with $\kappa-1$ degrees of freedom. The significance level $\alpha$ specifies the sensitivity of the test.

\subsection{Fast Discord Discovery} \label{sec:HOTSAX}
A time series discord of length $l$ is defined as the subsequence with the highest distance from its nearest neighbour subsequence of the same length. To exclude trivial matches, only non-overlapping subsequences are considered in the nearest neighbour search. Discords have been proposed in~\cite{bib:HOTSAX} as a consistent metric to describe anomalies.

The time complexity of the brute force method for discord discovery, namely, comparing all pairs of non-overlapping subsequences, is at the order of $O(N^2)$, where $N$ is the length of the time series. However, the heuristic HOT-SAX algorithm allows early stopping the searching loop, producing identical results yet in 3 or 4 orders of magnitude speedup~\cite{bib:HOTSAX}. The HOT-SAX algorithm leverages the SAX representation of the subsequences to quickly order them from the most rare to the most common, simply by counting the subsequences with identical SAX representation. A detailed description of the HOT-SAX algorithm can be found in~\cite{bib:HOTSAX}. Note that since a detailed complexity analysis of the algorithm is not feasible, the running time of HOT-SAX is measured by the times the distance function is called.

\section{Statistical and Information-Theoretic Issues of SAX}\label{sec:information}

\subsection{The Gaussian Assumption}\label{sec:gaussian}

Although the Gaussian model is commonly used to describe the statistics in a broad range of data sources due to its theoretical justification and computational tractability, the normality assumption is often violated in various practical applications. As such, a mismatch in the distribution yields an information loss that can be quantified by the Kullback-Leibler divergence. As we analyze in Sec.~\ref{sec:psax_exp}, this information loss degrades the performance of SAX.

In the following, we quantify the information that is lost during data source quantization due to a mismatch between the Gaussian distribution assumption and the actual underlying data distribution. The amount of lost information is described by the increment in the average number of bits required to describe a sample with a specific precision, when the assumed distribution is wrong. To do so, first we introduce the differential entropy and the Kullback-Leibler divergence of continuous random variables.

The differential entropy $h(f_X)$ of a continuous random variable $X$ with pdf $f_X(x)$ is defined by
\begin{equation}\label{eq:diff_entropy}
    h(f_X) \triangleq -\int f_X(x)\log{f_X(x)}dx\ ,
\end{equation}
where the integration is performed in the regions of the sample space where $f_X(x)>0$. Entropy is measured in \textit{bits} if the $\log$ base is $2$ and in \textit{nats} if it is $e$.

The Kullback-Leibler divergence $D_{KL}(f_X\!\parallel\! f_G)$ between the probability densities $f_X(x)$, $f_G(x)$ (in this order) is defined by
\begin{equation}
    D_{KL}(f_X \!\parallel\! f_G) \triangleq \int{f_X(x)\log{\frac{f_X(x)}{f_G(x)}}dx}\ ,
\end{equation}
with the convention that $0\log{0}=0$.
Notice that
\begin{equation}\label{eq:KL_cont_ineq}
    D_{KL}(f_X \parallel f_G)\geq 0 \ ,
\end{equation}
with equality iff $f_X(x)=f_G(x)$ almost everywhere. Furthermore,
\begin{equation}\label{eq:KL_cont_cross}
    D_{KL}(f_X \parallel f_G) = h(f_X,f_G)-h(f_X)\ , 
\end{equation}
where $h(f_X,f_G)\triangleq\int{f_X(x)\log{f_G(x)}dx}$ is the continuous cross-entropy.

Let $X^{\Delta}$ denote the quantized version of $X$, defined by
\begin{equation}
    X^{\Delta} = x_i\ ,\enskip \mathrm{if}\enskip i\Delta \leq X < (i+1)\Delta\ .
\end{equation}
Then, the probability mass function (pmf) of $X^{\Delta}$ is given by~\cite[Sec. 8.3]{bib:Thomas}:
\begin{equation}\label{eq:quantized_X_pmf}
    P(x_i) =  \int_{i\Delta}^{(i+1)\Delta}f_X(x)dx\ .
\end{equation}

For convenience in calculations, the above expressions rely on quantization intervals of equal length $\Delta$. However, for any quantization scheme it holds that, as the number of the quantization intervals approaches infinity, they become equally infinitesimal.

The next theorem provides upper and lower bounds for the expected description length of $X^{\Delta}$ and is the continuous analog of~\cite[Thm. 5.4.3]{bib:Thomas}.
\begin{theorem}\label{thm:wrong_code_continuous}
The expected description length, $\mathbb{E}[l(X^{\Delta})]$, of the random variable $X^{\Delta}$, assuming optimal coding under the probability density function $f_G(x)$, is bounded as follows,

\begin{align}
    \mathbb{E}[l(X^{\Delta})] & \geq h(f_X)+D_{KL}(f_X \!\parallel\! f_G)+\log{\frac{1}{\Delta}} \ , \nonumber\\
    \mathbb{E}[l(X^{\Delta})] & < h(f_X)+D_{KL}(f_X\! \parallel\! f_G)+\log{\frac{1}{\Delta}}+1 \ .
\end{align}
\end{theorem}

The detailed proof is derived in the Appendix. Notice that the additive term $\log{\frac{1}{\Delta}}$ is the contribution of the quantization. Theorem~\ref{thm:wrong_code_continuous} together with \eqref{eq:KL_cont_ineq} denote that the information loss by assuming that the distribution of $X$ is equal to $f_G(x)$ instead of $f_X(x)$ is given by $D_{KL}(f_X\! \parallel\! f_G)$. This is similar to the result derived in the case of discrete random variables. When $f_G(x)$ is the Gaussian density function, the following corollary holds.

\begin{corollary}
Let $X$ be an arbitrary continuous random variable with density $f_X(x)$, with mean $\mu_X$ and variance $\sigma_X^2$, and $G$ a Gaussian random variable, $G\sim \N(\mu_G,\sigma_G^2)$, whose density is denoted by $f_G(x)$. Then,
\begin{equation}\label{eq:info_loss}
    D_{KL}(f_X\! \parallel\! f_G)\! =\! \ln\sigma_G\sqrt{2\pi} +\frac{1}{2\sigma_G^2}\int f_X(x) (x-\mu_G)^2 dx - h(f_X),
\end{equation}
measured in nats.
\end{corollary}

\begin{proof}
\begin{equation}\label{eq:coroll_eq1}
\begin{aligned}
    h(f_X,f_G)  =& -\int f_X(x) \ln f_G(x)dx\\
            =& -\int f(x) \ln\frac{1}{\sqrt{2\pi\sigma_G^2}}e^{-(x-\mu_G)^2/(2\sigma_G^2)}dx\\
            =& -\int f_X(x) \left(\ln\frac{1}{\sqrt{2\pi\sigma_G^2}} - \frac{(x-\mu_G)^2}{2\sigma_G^2}\right)dx\\
            =& -\int f_X(x) \ln\frac{1}{\sqrt{2\pi\sigma_G^2}}dx + \int f_X(x) \frac{(x-\mu_G)^2}{2\sigma_G^2}dx\\
            =& \ln \sigma_G\sqrt{2\pi} +\frac{1}{2\sigma_G^2}\int f_X(x) (x-\mu_G)^2 dx\ .
\end{aligned}
\end{equation}
Combining~\eqref{eq:KL_cont_cross} and~\eqref{eq:coroll_eq1} completes the proof.
\end{proof}

Under the constraints that $\mu_X=\mu_G=0$ and $\sigma_X=\sigma_G=1$, which is the case in SAX due to Z-normalization, ~\eqref{eq:info_loss} reduces to
\begin{equation}\label{eq:sax_loss}
\begin{aligned}
    D_{KL}(f_X\! \parallel\! f_G) &= \ln \sqrt{2\pi} +\frac{1}{2}\int f_X(x) x^2 dx - h(f_X)\\
                          &= \ln \sqrt{2\pi}+\frac{1}{2}\cdot 1 - h(f_X)\\
                          &= \ln \sqrt{2\pi e} - h(f_X)\quad \mathrm{nats}\\
                          &= \frac{1}{\ln 2}\left(\ln \sqrt{2\pi e} - h(f_X)\right)\quad \mathrm{bits.}
\end{aligned}
\end{equation}
Eq.~\eqref{eq:sax_loss} quantifies the information loss in the discretization step of SAX assuming Z-normalized PAA samples.

\subsection{Z-normalization and Piecewise Aggregate Approximation}\label{sec:Z-norm_PAA}

This section elaborates on a fundamental misunderstanding when applying the conventional SAX method. 
Specifically, even when the Gaussian assumption holds for a given data set, performing a Z-normalization on the input data does not guarantee a standard distribution for the PAA sequence.

To verify this statement, consider the case where a time series $X\in \T^N$ is a sequence of correlated and jointly Gaussian-distributed random variables. We set $M=N/m$, that is, the PAA transformation $X \rightarrow Y$ is given by $y_{i} = \left[ \sum_{k=m(i-1)+1}^{mi} x_{k}\right]/m,\enskip i=1,\dots,M$. We assume that the samples are Z-normalized, thus $x_k\sim \N(0,1)\ \forall\, k=1,\dots,N$ and we denote by $\bar{\rho}$ the average pairwise Pearson's correlation of the distinct sample pairs. Then, it can be derived in a straightforward manner that $y_{i} \sim \N(\mu_Y,\sigma_Y^2)$, where
\begin{equation}
\mu_Y = 0\ , \quad \sigma_Y^2 = \frac{1+ (m-1)\bar{\rho}}{m}\ .
\end{equation}
That is, the PAA sequence $Y$ follows a standard Gaussian distribution only when $\bar{\rho}=1$.

The above theoretical proof explains the experimental results described in~\cite{bib:SAXequiprob}, where the authors observe that the PAA step reduces the resulting standard deviation.

Notice that in Sec.~\ref{sec:gaussian} the PAA sequence was assumed to have a unit variance. Nevertheless, if Z-normalization is performed prior to the PAA step, then, $\sigma_Y^2\leq 1$ as showed above, where the strict inequality holds for $\bar{\rho}<1$. In this case, the information loss~\eqref{eq:info_loss} in the discretization step is equal to
\begin{equation}\label{eq:sax_loss_PAA}
    D_{KL}(f_Y\! \parallel\! f_G) = \frac{1}{\ln 2}\left(\ln\sqrt{2\pi} + \frac{1}{2}\sigma_Y^2 - h(f_Y) \right)\quad \mathrm{bits,}
\end{equation}
where $f_Y(y)$ is the pdf of $Y$. By noticing that the effect of the standard deviation on the differential entropy is an additive $log$ term, the right side of~\eqref{eq:sax_loss_PAA} is larger than the one of~\eqref{eq:sax_loss} by $\left((\sigma_Y^2-1)/2-\ln\sigma_Y\right)/\ln 2$.

Based on the above derivations and remarks, it becomes clear that there exists an inherent mistake in the default SAX procedure that induces an additional information loss. This is due to the fact that the PAA sequence is assumed to have unit standard deviation, which is not true.

To address this issue, we propose a simple, yet effective, solution. Specifically, instead of performing a Z-normalization on the original data samples, the PAA samples are Z-normalized directly which also guarantees their unit variance.

\section{Data-Driven SAX-based Symbolic Representations}\label{sec:represent_prelim}

Motivated by the remarks in the previous section, we introduce a new family of SAX-based symbolic representations. We enable increased flexibility in utilizing appropriate discretization schemes depending on the desired application scenario. The proposed representations adapt directly to the inherent data statistics, minimizing the information loss in the discretization step. In particular, as proved in Sec.~\ref{sec:gaussian}, the information loss in terms of \textit{average coding length} is minimized with the Kullback-Leibler divergence between the true and the assumed distribution. Thus optimal coding length is achieved by estimating the true distribution prior to encoding. At this point notice that the Kullback-Leibler divergence (a.k.a. \textit{relative entropy}) belongs to the Shannon-type measures of information entropy. 
Another type of entropy, namely Renyi’s entropy, is a generalization of Shannon's entropy and a certain form of it, \textit{Renyi's cross-entropy}, has been linked with information-theoretic learning. Specifically, the authors in~\cite{bib:MS_information} prove that mean-shift clustering minimizes Renyi's cross entropy, which implies minimization of the associated (Renyi's) information loss.

We exploit the above observations to develop two novel SAX-based representations. In particular, our representations rely on density-based discretization methods, namely the Lloyd-Max quantization and the mean-shift clustering algorithms. 
The following subsections overview the key components of our proposed symbolic representations.

\subsection{Kernel Density Estimation}\label{sec:KDE}

Kernel density estimation (KDE)~\cite{bib:KDE} is a method for estimating the probability density function of a data source. Essentially, it is the summation of a set of translated and dilated kernel functions centered at each observed sample,
\begin{equation}\label{eq:KDE}
    \hat{f}_{h,K}(x) = \frac{1}{Nh}\sum_{i=1}^N K\left(\frac{x-x_i}{h}\right)\ ,
\end{equation}
where $x_i$, $i=1,\ldots,N$, are the observed samples, $K(\cdot)$ is the kernel function that controls the weight given to the neighboring samples of $x\in X$, and $h$ is a smoothness parameter, which controls the width of the kernel. In general, the smoothness parameter $h$ affects more the accuracy of the estimator than the kernel function itself.

\begin{figure}
    \centering
    \subfloat[Epanechnikov kernel]{\includegraphics[width=.49\linewidth]{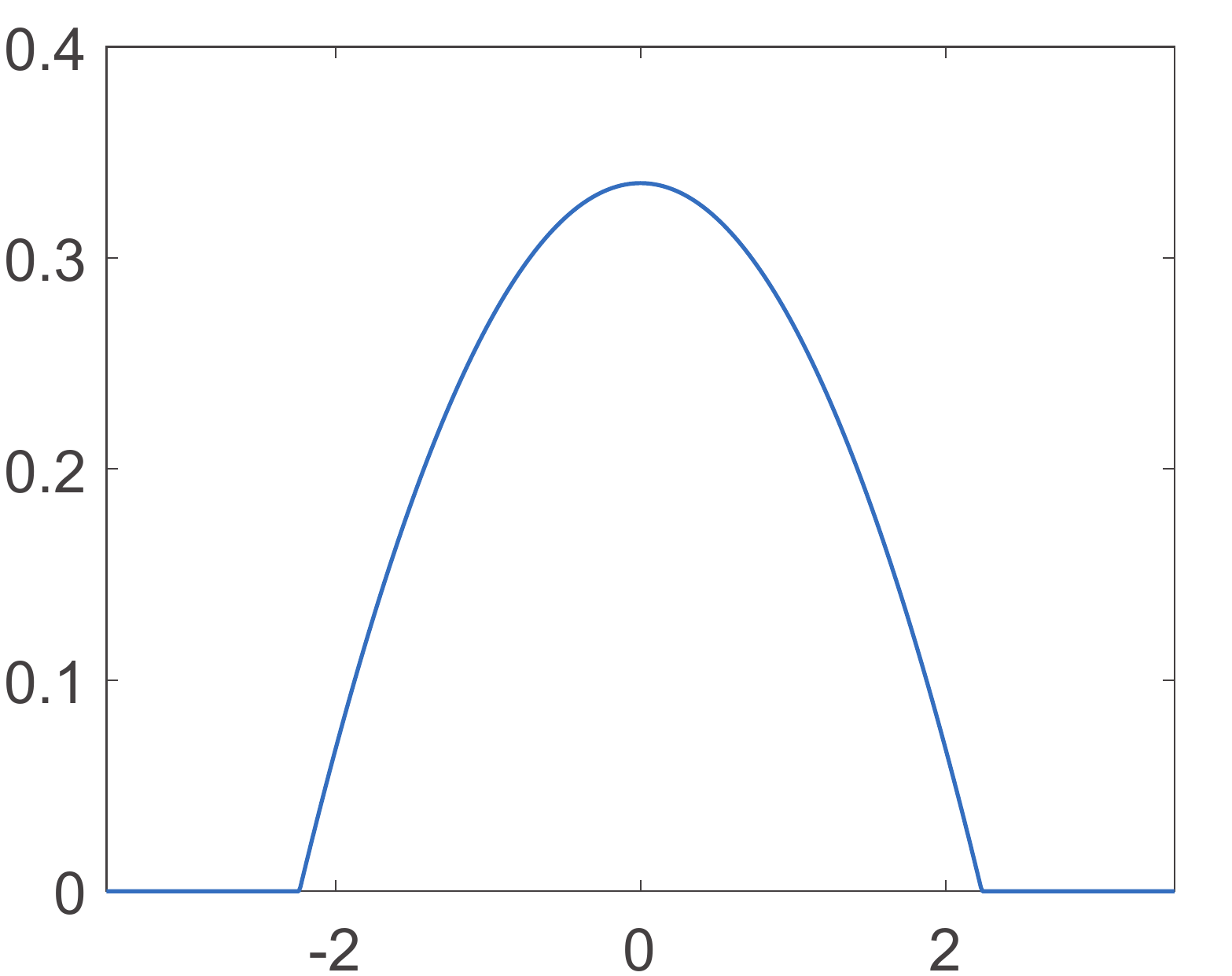}%
    \label{fig:Epanech}}
    \hfill
    \subfloat[Gaussian Kernel]{\includegraphics[width=.49\linewidth]{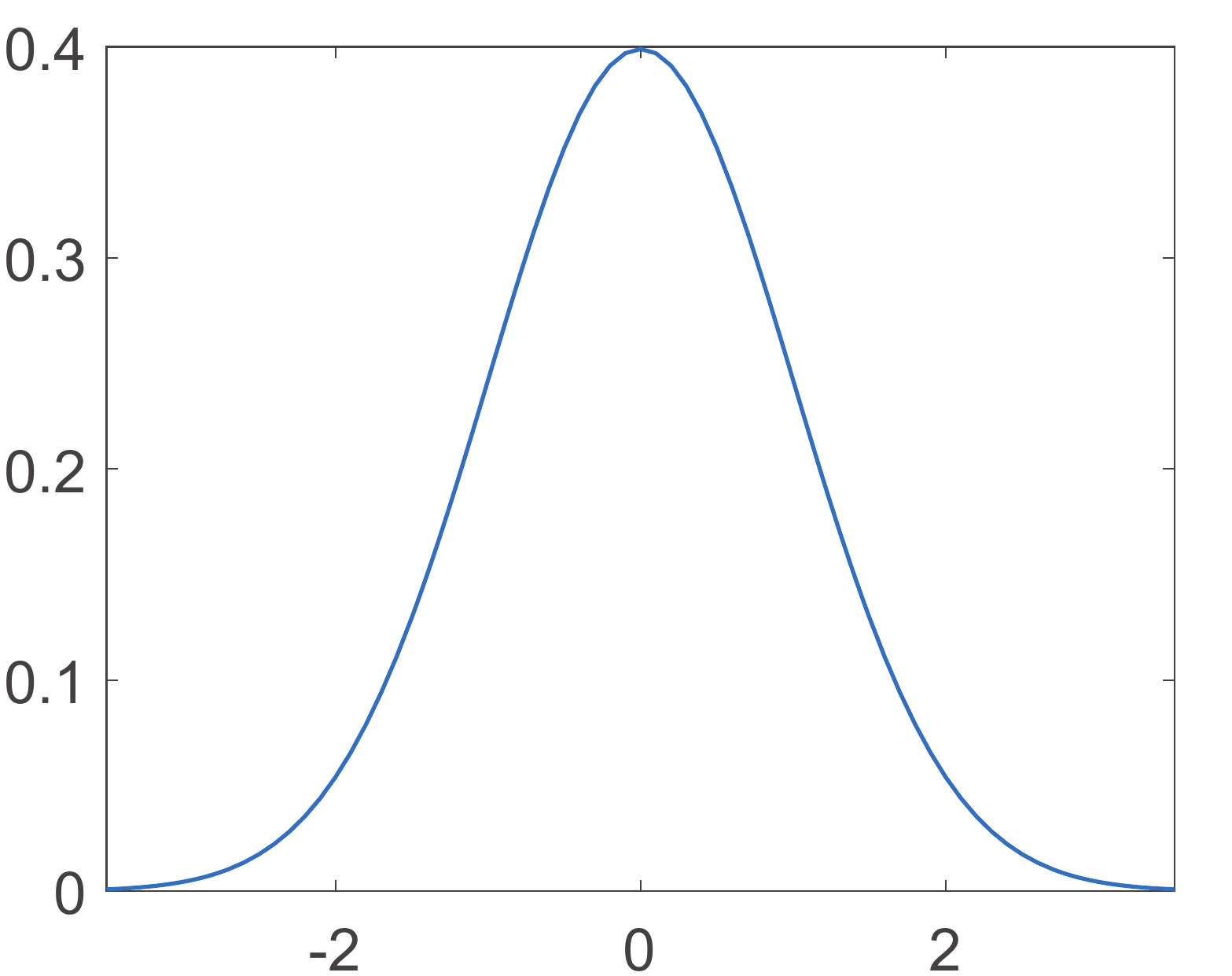}%
    \label{fig:Gaussian}}
    \caption{The kernel functions employed by our proposed method for density estimation.}
\end{figure}

Under suitable conditions, the optimal kernel, in terms of the asymptotic minimum integrated squared error (AMISE) between the estimated and the target density function, is proved to be the Epanechnikov kernel~\cite{bib:Epanechnikov}, shown in Fig.~\ref{fig:Epanech}. Another commonly used kernel is the Gaussian kernel, shown in Fig.~\ref{fig:Gaussian}. These two kernels are defined by:
\begin{align}
    &\text{Epanechnikov:}&& K(x)=\max\left(0,\frac{3}{4\sqrt{5}}(1-\frac{x^2}{5})\right) \label{eq:epanechnikov}\\
    &\text{Gaussian:}&& K(x)=\frac{1}{\sqrt{2\pi}}e^{-\frac{1}{2}x^2}\label{eq:gaussian}
\end{align}

For a given kernel function, the optimal value for $h$, in terms of a minimum AMISE, is given by
\begin{equation}\label{eq:hopt}
    h_{opt} = \left[\frac{\int_{-\infty}^{+\infty}K^2(y)dy}{N \mu_2^2\int_{-\infty}^{+\infty}(f^{''}(x))^2dx}\right]^{1/5}\ ,
\end{equation}
where $K(x)$ is the chosen kernel function, $\mu_2$ is the second moment of $K(x)$, and $f(x)$ is the target pdf.

Since, in practice, the knowledge of the true target pdf is typically not available, the authors in~\cite{bib:Silverman} suggested that a known family of distributions can be used to calculate $f^{''}(x)$. The Gaussian is the most commonly used distribution, which yields the following approximation for the optimal smoothness parameter,
\begin{equation}\label{eq:hoptN}
    h_{opt}^{\N} = \hat{\sigma} \cdot \left[\frac{8\sqrt{\pi}\int_{-\infty}^{+\infty}K^2(y)dy}{3 N \mu_2^2}\right]^{1/5}\ ,
\end{equation}
where $\hat{\sigma}$ is the estimated standard deviation of the data.

For the Epanechnikov and Gaussian kernels, the following values are replaced in~\eqref{eq:hoptN},
\begin{align}
    &\text{Epanechnikov:}&&\int_{-\infty}^{+\infty}K^2(y)dy = \frac{3}{5}\ ,& \mu_2 = \frac{1}{5}\label{eq:Epanech_kernel_params}\\
    &\text{Gaussian:}&&\int_{-\infty}^{+\infty}K^2(y)dy = \frac{1}{2\sqrt{\pi}}\ ,& \mu_2 = 1\label{eq:Gauss_kernel_params}
\end{align}
yielding the following optimal smoothness parameters,
\begin{align}
    &\text{Epanechnikov:}&& h_{opt}^{\N}=2.3449\cdot\hat{\sigma}\cdot N^{-1/5}\label{eq:Epanech_hopt}\\
    &\text{Gaussian:}&& h_{opt}^{\N}=1.0492\cdot\hat{\sigma}\cdot N^{-1/5}\label{eq:Gauss_hopt}
\end{align}
The expressions in~\eqref{eq:Epanech_hopt}, \eqref{eq:Gauss_hopt} are referred as  Silverman's rules of thumb.

If some wishes to estimate the derivative of a density function, rather than the function itself, then the AMISE analysis yields different optimal parameters. Particularly, similar calculations as before give the following results:
\begin{align}
    &\text{Epanechnikov kernel:}&& h_{\nabla,opt}^{\N}=1.5232\cdot\hat{\sigma}\cdot N^{-1/7}\label{eq:Epanech_grad_hopt}\\
    &\text{Gaussian kernel:}&& h_{\nabla,opt}^{\N}=0.9686\cdot\hat{\sigma}\cdot N^{-1/7}\label{eq:Gauss_grad_hopt}
\end{align}
The latter result shall be useful for estimating the modes of the density. In the proposed framework, we employ both the Epanechnikov and the Gaussian kernel, with the choice of the kernel function depending on the specific discretization scheme that follows.

\subsection{Discretization Schemes}

\begin{figure}
    \centering
    \subfloat[Equiprobable quantization (entropy maximizer), $\kappa=8$.]{\includegraphics[width=\linewidth]{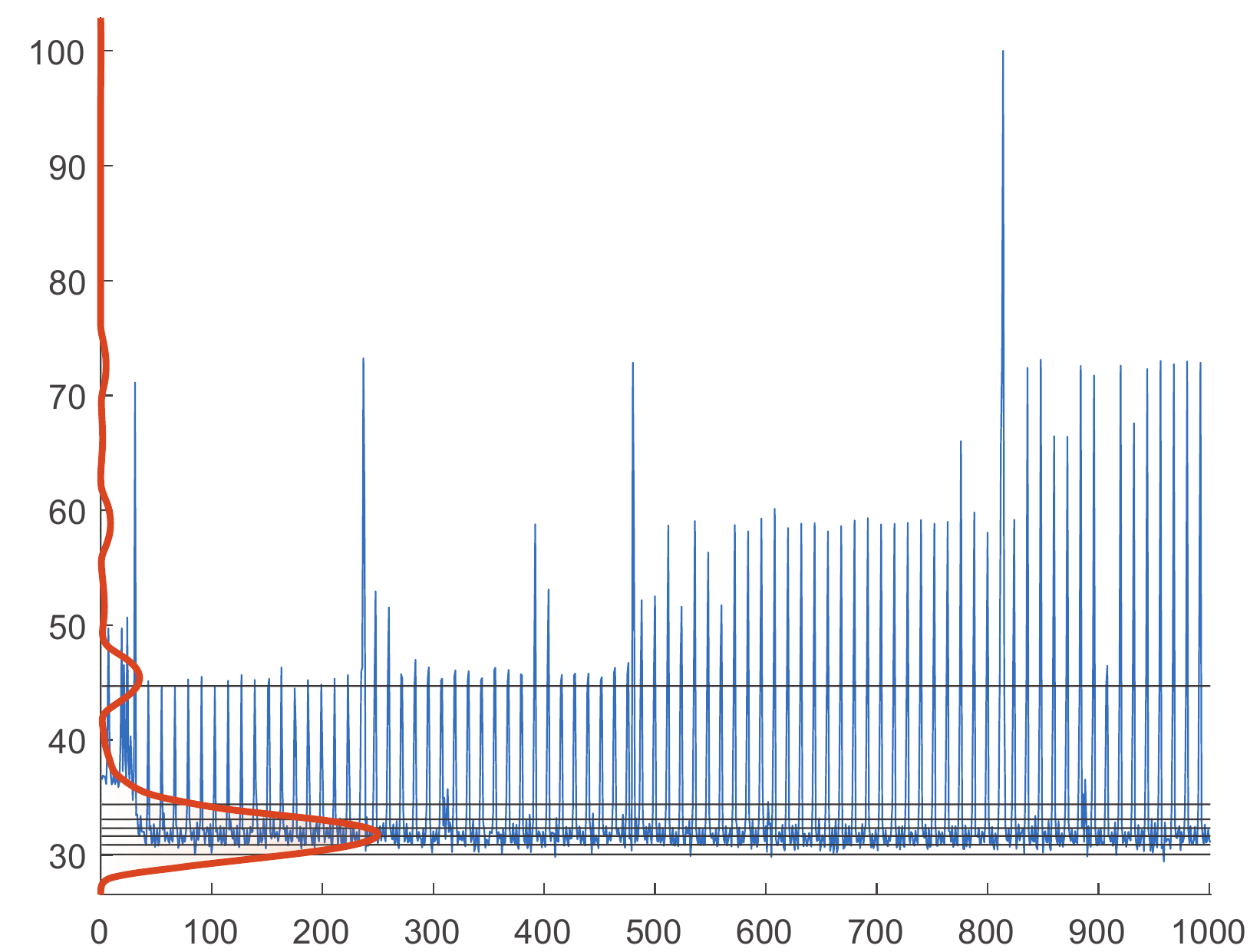}\label{fig:equi_quant}}
    \vspace{6pt}
    \subfloat[Lloyd-Max quantization (MSE minimizer), $\kappa=8$.]{\includegraphics[width=\linewidth]{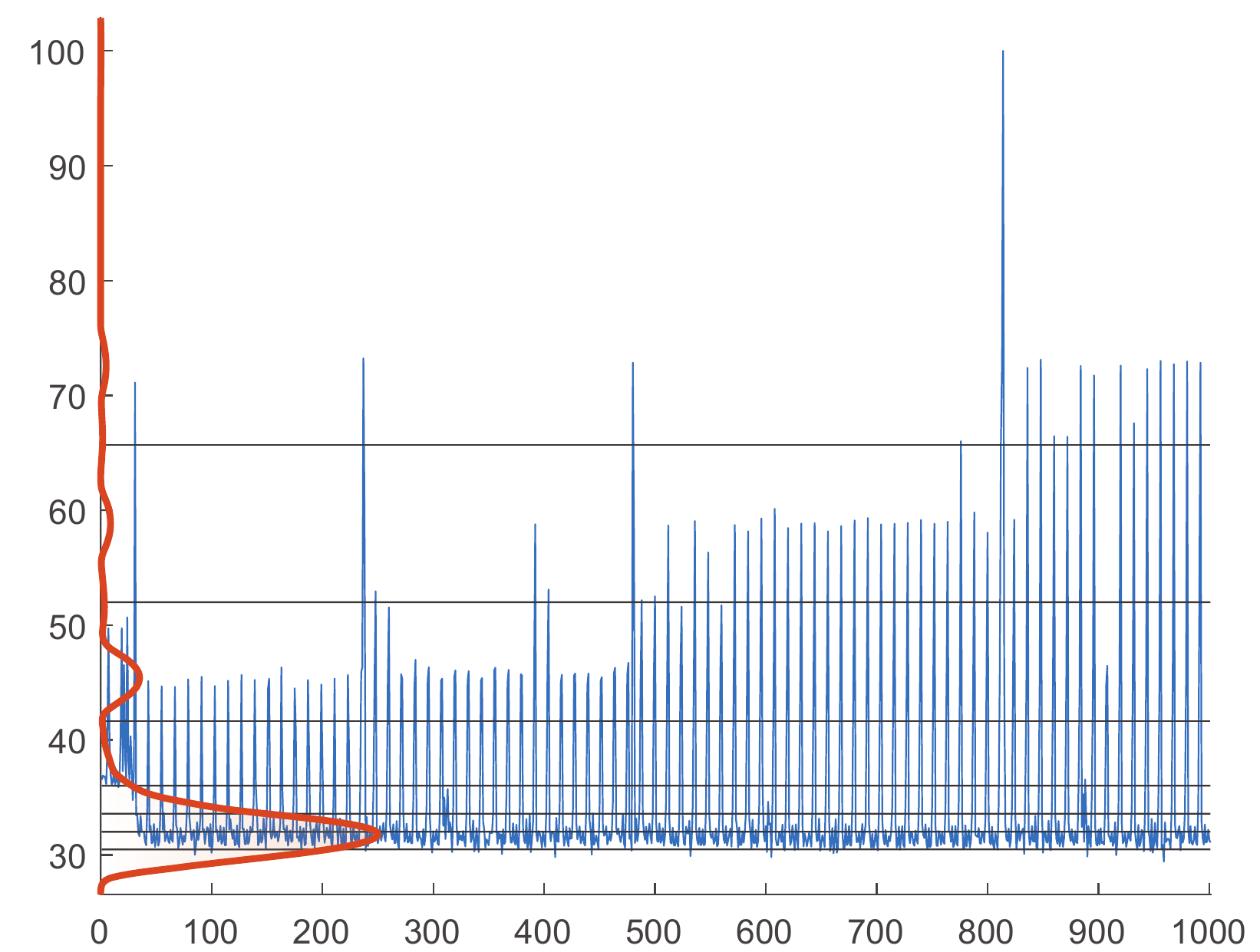}\label{fig:lloyd-max}}
    \vspace{6pt}
    \subfloat[Mean-shift clustering, $\kappa=9$ (automatically).]{\includegraphics[width=\linewidth]{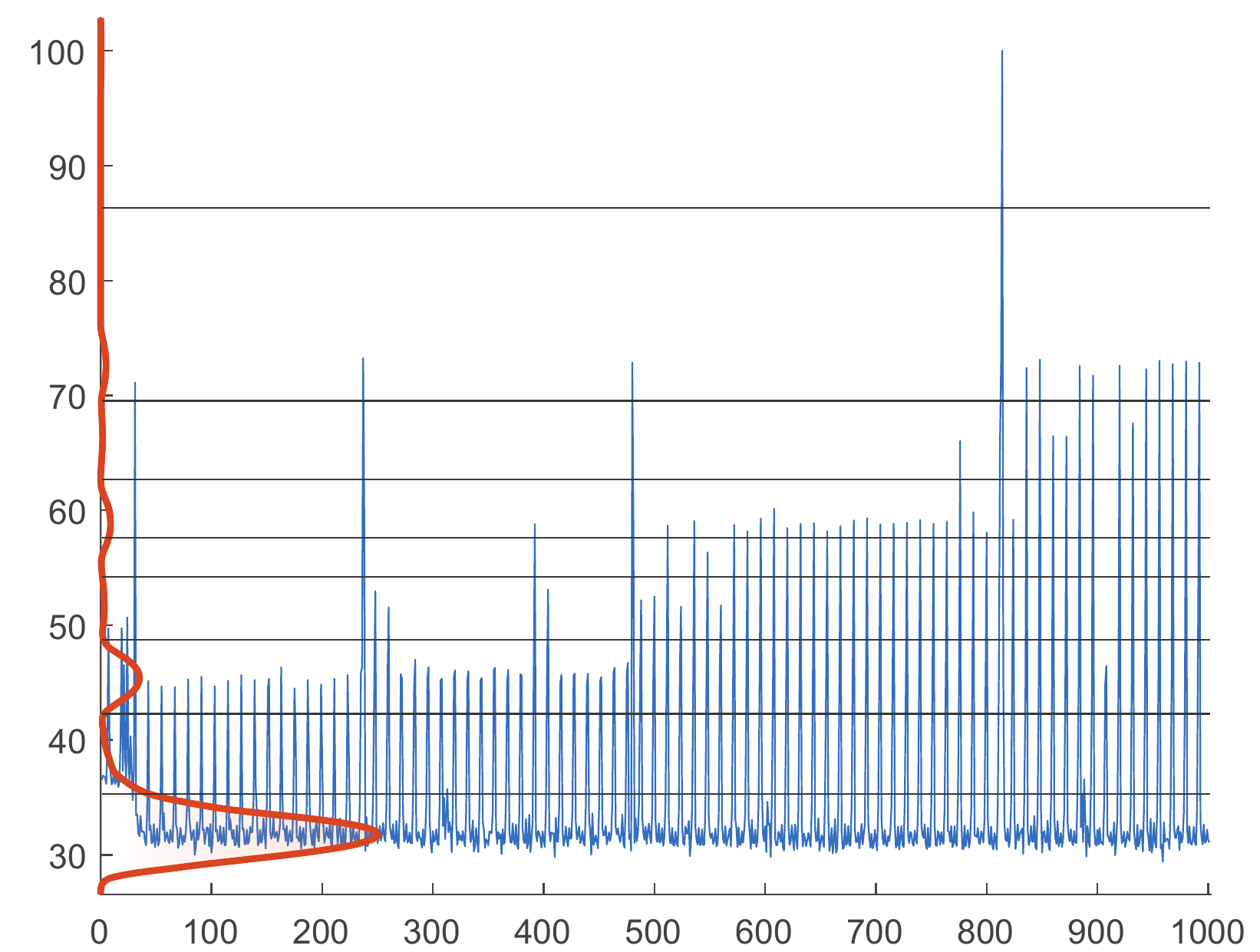}\label{fig:mean-shift}}
    \vspace{3pt}
    \caption{Various density-based discretization schemes. The density is estimated via KDE. Conventional SAX employs equiprobable quantization assuming a Gaussian distribution (ref. Fig.~\ref{fig:SAX}). Our pSAX employs the Lloyd-Max quantizer and cSAX employs the mean-shift clustering. Notice the decreasing amount of energy in the dominant mode.}
\end{figure}

Two distinct methods are employed in the discretization step, namely, (i) the Lloyd-Max quantizer and (ii) the mean-shift clustering algorithm. A visual comparison of the different data-driven discretization schemes is depicted in Figs.~\ref{fig:equi_quant}-\ref{fig:mean-shift}.

\subsubsection{Lloyd-Max Quantization}\label{sec:Lloyd-Max}

The Lloyd-Max quantizer~\cite{bib:Lloyd} is an alternating MSE minimization algorithm.
Let $\{X_1,\dots,X_\kappa\}$ be a partition of the sample space $\X$, where $X_i$'s are disjoint subsets with an arbitrary number of input elements from $\X$, and $\{c_1,\dots,c_\kappa\}$ be a set of codewords. Then, the quantization mapping $Q(\cdot)$ is defined by
\begin{equation}
    Q(x_j) = c_i\ ,\quad \forall~x_j\in X_i\ .
\end{equation}
The number of quantization intervals, $\kappa$, is a predefined parameter of the quantization process.

Let $d(x-Q(x))$ denote a distance measure between an input element and the codeword assigned to it. If $f(x)$ is the underlying probability density function of the input space, and $b_i$, $b_{i+1}$ are the infimum and supremum of the subset $X_i$, respectively, then, the distortion $D$ is defined as the expected value of $d$, that is,
\begin{equation}
    D = \mathbb{E}_f[d(x-Q(x))] = \sum_{i=1}^\kappa \int_{b_i}^{b_{i+1}}d(x-c_i)f(x)dx\ .
\end{equation}

The necessary criteria for minimizing the distortion function $D$ are derived by calculating the partial derivatives with respect to $b_i$'s and $c_i$'s and setting them equal to zero. In the special case where the distance measure $d$ is the Euclidean distance, the computations yield the following necessary criteria:
\begin{equation}\label{eq:quant_Eucl_crit1}
   b_i = (c_i+c_{i-1})/2\ ,\quad i=2,\dots,\kappa
\end{equation}
\begin{equation}\label{eq:quant_Eucl_crit2}
    c_i = \frac{\int_{b_i}^{b_{i+1}}xf(x)dx}{\int_{b_i}^{b_{i+1}}f(x)dx}\ ,\quad i=1,\dots,n
\end{equation}
That is, $b_i$ must be equal to the midpoint of $c_i$ and $c_{i-1}$, whilst $c_i$ must be the centroid of subset $X_i$. Note that the above criteria are necessary but not sufficient. In order to be sufficient, the Hessian matrix of the distortion function must be positive definite.

Due to their mutual relationship, the solution of \eqref{eq:quant_Eucl_crit1} and \eqref{eq:quant_Eucl_crit2} is  approximated iteratively by alternating between the necessary criteria. The process is summarized in Algorithm~\ref{alg:Lloyd-Max}, while a visual example is illustrated in Fig.~\ref{fig:lloyd-max}.

~
\algcaption{Lloyd-Max Quantizer\\
\textbf{Input:} $f(x)$ (probability density function)}\label{alg:Lloyd-Max}
\begin{algorithmic}[1]
\State Initialize range: $b_1\gets -\infty$, $b_{\kappa+1}\gets +\infty$.
\State Initialize codewords randomly $c_i\in (b_1,b_{\kappa+1}),\ i=1,\dots,\kappa$.
\Repeat
\State $b_i\gets (c_i+c_{i-1})/2$ ,\quad for $i=2,\dots,\kappa$\label{alg:Lloyd-Max:cond1}
\State $c_i\gets \frac{\int_{b_{i}}^{b_{i+1}}xf(x)dx}{\int_{b_{i}}^{b_{i+1}}f(x)dx}$ ,\quad for $i=1,\dots,\kappa$ \label{alg:Lloyd-Max:cond2}
\Until{convergence}
\end{algorithmic}

\subsubsection{Mean-Shift Clustering}

Mean-shift clustering~\cite{bib:Mean-Shift02} is based on the rationale that the generated samples form clusters corresponding to regions with high local probability density. That is, the clusters correspond to the modes of the underlying probability density function. In the following, we describe the procedure for uni-dimensional signals, however, the generalization to higher-dimensional signals follows in a straightforward way.

Given the outcome of the KDE process~\eqref{eq:KDE}, the modes of the estimated density function $\hat{f}_{h,K}(x)$ are found next. The mean-shift approach estimates the modes by relying on kernels that are symmetric around zero and can be expressed as
\begin{equation}\label{eq:profile}
    K(x) = z_{k}k(\|x\|^2)\ ,
\end{equation}
where $z_{k}$ is a normalization constant parameter that enforces $K(x)$ to integrate to one. The function $k(x)$ is called the \textit{profile} of $K(x)$.

By employing~\eqref{eq:profile}, the estimator in~\eqref{eq:KDE} is given by
\begin{equation}
    \hat{f}_{h,K}(x) = \frac{z_k}{Nh}\sum_{i=1}^N k\left(\left\|\frac{x-x_i}{h}\right\|^2\right)\ .
\end{equation}

Then, the modes are found by detecting the points where the gradient $\nabla \hat{f}_{h,K}(x)$ equals zero. These points are the stationary points, including all the local maxima, that is, the modes of the density function $\hat{f}_{h,K}(x)$. The gradient is given by
\begin{equation}\label{eq:grad1}
    \nabla \hat{f}_{h,K}(x) = \frac{2z_k}{Nh^3}\sum_{i=1}^N (x-x_i) k'\left(\left\|\frac{x-x_i}{h}\right\|^2\right)\ .
\end{equation}

Defining the profile $g(x) = -k'(x)$ and the corresponding kernel $G(x) = z_g g(\|x\|^2)$, the gradient~\eqref{eq:grad1} equals 
\begin{equation}\label{eq:grad2}
\begin{aligned}
    \nabla \hat{f}_{h,K}(x) &= \frac{2z_g}{Nh^3}\sum_{i=1}^N (x_i-x) g\left(\left\|\frac{x_i-x}{h}\right\|^2\right)\\
                            = \frac{2z_g}{Nh^3}&\left[\sum_{i=1}^N x_i g\left(\left\|\frac{x_i-x}{h}\right\|^2\right) - x\sum_{i=1}^N g\left(\left\|\frac{x_i-x}{h}\right\|^2\right)\right]\\
                            = \frac{2z_g}{Nh^3}&\left[\sum_{i=1}^N g\left(\left\|\frac{x_i-x}{h}\right\|^2\right)\right]\left[\frac{\sum\limits_{i=1}^N x_i g\left(\left\|\frac{x_i-x}{h}\right\|^2\right)}{\sum\limits_{i=1}^N g\left(\left\|\frac{x_i-x}{h}\right\|^2\right)}-x\right]\\
\end{aligned}
\end{equation}
The second term of the last equation in~\eqref{eq:grad2} is called the \textit{mean-shift} vector $m_{h,g}$,
\begin{equation}\label{eq:mean-shift}
    m_{h,g} \triangleq \frac{\sum\limits_{i=1}^N x_i g\left(\left\|\frac{x_i-x}{h}\right\|^2\right)}{\sum\limits_{i=1}^N g\left(\left\|\frac{x_i-x}{h}\right\|^2\right)}-x\ ,
\end{equation}
which can be rewritten in terms of the estimated density function as follows,
\begin{equation}\label{eq:mean-shift2}
    m_{h,g} = \frac{Nh^2}{2z_k}\cdot\frac{\nabla \hat{f}_{h,K}(x)}{\hat{f}_{h,G}(x)}\ .
\end{equation}

The above equation implies that the mean-shift vector points at the same direction as the gradient of the estimated density $\hat{f}_{h,K}(x)$, which is the direction with the highest increase in density. Furthermore, it is inversely proportional to $\hat{f}_{h,G}(x)$, which affects the mean-shift vector by decreasing its value in the areas with high density, that is, near the modes of the density function. These remarks lead to the formulation of the mean-shift procedure, which is essentially a gradient ascend algorithm with a self-adjusted step-size. The mean-shift clustering algorithm is summarized in Algorithm~\ref{alg:Mean-shift}, while a visual example is illustrated in Fig.~\ref{fig:mean-shift}.

~
\algcaption{Mean-Shift Clustering}\label{alg:Mean-shift}
\begin{algorithmic}[1]
\While{stopping criteria are not met}
\State Pick a random sample $x$.
\Repeat
\State Compute the mean-shift $m_{h,g}(x)$.
\State Move to the next sample in the direction and step-size of $m_{h,g}(x)$.
\Until{convergence}
\EndWhile
\end{algorithmic}
~

In the mean-shift nomenclature, $K(x)$ is called the \textit{shadow} of $G(x)$. The Gaussian kernel is the only kernel that is the shadow of itself.

\section{Data-Driven SAX With Tight Lower Bounding Distances}\label{sec:psax}

This section analyzes our proposed data-driven symbolic representation, hereafter called probabilistic SAX (pSAX), which achieves tighter lower-bounding distances than the previous SAX-based counterparts. Furthermore, our method performs in a completely distribution agnostic framework, without any prior assumption for the data distribution. As such, it is an improved alternative to the conventional SAX representations for general data mining and indexing scenarios, where the data generating process is often unknown and the tightness of lower bound (TLB) (ref.~\eqref{eq:TLB}) is of utmost importance.

To this end, our method applies a kernel density estimator (KDE) (ref. Sec.~\ref{sec:KDE}) directly after the PAA step, in order to estimate the underlying probability density function of the PAA sequence, without any prior probabilistic assumption. Subsequently, the KDE step is coupled with a Lloyd-Max quantizer, as described in Sec.~\ref{sec:Lloyd-Max}, to estimate the optimal quantization intervals.

Notice also that, since our method does not rely on any prior assumption on the distribution of either the data or PAA, the $Z$-normalization step is needless. Nevertheless, in light of the observations made in Sec.~\ref{sec:Z-norm_PAA}, we estimate the density function of the PAA sequence, instead of the original time series itself.

\subsection{Implementation Details}

The KDE, which is the first step of the discretization process, is critical for guaranteeing the subsequent optimal quantization by means of Lloyd-Max (ref. Alg.~\ref{alg:Lloyd-Max}). In our implementation, the Epanechnikov kernel~\eqref{eq:epanechnikov} is employed for minimizing the AMISE criterion, and the smoothness parameter is set according to Silverman's rule in~\eqref{eq:Epanech_hopt}.

In order to improve the convergence rate and accuracy of the quantizer, the codewords are initialized by using the $k$-means$++$ algorithm~\cite{bib:k-meanspp}, which employs the raw PAA samples for the fast calculation of a good starting point before optimizing upon the estimated density function. This yields an improved performance, in terms of both convergence speed and a lower MSE.

\subsection{Novel Distance Measures}

Notice that the $mindist$ and $mindist\_PAA$ measures, defined in~\eqref{eq:mindist} and~\eqref{eq:mindist_PAA}, respectively, are lower bounds of the true Euclidean distance regardless of the chosen quantization intervals, and hence the lower-bounding property still holds in our proposed representation.

On the other hand, the conventional SAX does not provide arithmetic values for the symbols in the lower-dimensional space, as opposed to the Lloyd-Max quantizer, which is able to do so. As discussed in Sec.~\ref{sec:Lloyd-Max}, these values are precisely the centroids of the quantization intervals. This feature can be further exploited to define a new distance measure between two symbolic sequences $Q,\,C\in\C^M_A$, as follows,
\begin{equation}\label{eq:novel_dist_SAX}
    d_{s}(Q,C) = \sqrt{\frac{N}{M}\sum_{i=1}^M \left(q_i-c_i\right)^2}\ .
\end{equation}

Although the above distance measure does not lower bound the Euclidean distance, it is the closest to the true one under an MSE criterion. This proximity is up to a distortion caused by (i) the difference between the KDE-based and the true density and (ii) the incapability of Lloyd-Max quantization to minimize globally the MSE. Accordingly, a distance measure between a time series and a symbolic sequence can be derived, which is not possible in the conventional SAX case. Specifically, given $U\in\T^N$ and $C\in\C^M_A$, their distance is defined by
\begin{equation}\label{eq:error}
    d_{e}(U,C) = \sqrt{\frac{1}{N}\sum_{i=1}^M \left(\sum_{j=(N/M)(i-1)+1}^{(N/M)i}\left(u_j-c_i\right)^2\right)}\ .
\end{equation}
In the special case where $C$ is the symbolic representation of $U$, the $d_{e}$ is the root mean squared error (RMSE) between $U$ and its reconstruction from $C$.

Note that the proposed method does not need the $Z$-normalization pre-processing step. Nevertheless, in order to ensure that the RMSE has the same scale for all representations, we $Z$-normalize all datasets before evaluating the methods.

\section{Data-Driven SAX for Efficient High-Level Time Series Mining}\label{sec:csax}

This section describes a novel SAX representation, hereafter called clustering SAX (cSAX), that provides a descriptive representation for high-level data mining tasks, such as anomaly detection, which is investigated in this work. Although our symbolic representation is designed primarily for data mining tasks that benefit from density-based clustering, it still holds the attractive properties of SAX for database indexing, thus it can be easily integrated in relative systems.

The noticeable difference that makes our proposed symbolic representation effective for anomaly detection lies on the choice of the mean-shift clustering (ref. Alg.~\ref{alg:Mean-shift}) for the discretization step. Notice that the alphabet size in our symbolic representation is set automatically by the mean-shift algorithm, thus the only parameter to be defined for the symbolic transformation is the dimensionality $M$ of the symbolic representation.

\subsection{Implementation Details}

For the mean-shift clustering process we select the Gaussian kernel, instead of the AMISE-optimal Epanechnikov kernel. This choice is motivated by the fact that the Epanechnikov kernel and its derivative are non-continuous at $\pm\sqrt{5}$, which is undesirable for mode estimation. The smoothness parameter is set according to~\eqref{eq:Gauss_grad_hopt}, whilst the window length $n$, which is the only hyper-parameter of our anomaly detection method, is fixed empirically to $50$ for all the datasets.

A problem that arises in practical applications is the need for a training phase of the mean-shift algorithm. The potential lack, or the availability of a very limited amount, of historical data from a given data source makes this problem even more complicated. In that case, the clustering method should be able to adapt on-the-fly to the evolving time series stream. To address this problem, a suitable criterion for dynamically re-estimating the clusters from the streaming data is introduced in the next subsection.

\subsection{Dynamic Mean-Shift Clustering}\label{sec:dynamic_MS}

Having estimated an initial set of clusters, a criterion for dynamically re-estimating the clusters is defined, by utilizing all the data up to the current data point. There are two extreme cases: (i) the clusters are always re-estimated, and (ii) the clusters are never re-estimated. In the first case, the clusters are always up to date, but with a severe delay each time a sample arrives, whilst in the latter case there is no delay, but the estimation is always out of date. None of the above extremes is desirable in practice, thus we propose a computationally tractable process for lying in between the two extremes.

Consider that the current clusters have been estimated at some time using the mean-shift vector in~\eqref{eq:mean-shift2} with an arbitrary kernel $G$, smoothness parameter $h_G$, with $N$ samples. The mean-shift procedure implicitly estimates the density function by using another kernel $K$ that is the shadow of $G$. The kernels that are employed by our method are valid probability density functions, that is, they are non-negative and integrate to one. Hence, we can define the standard deviation $\sigma_K$ of the random variable whose pdf is equal to the kernel $K$. The proposed criterion is the following: 

\begin{definition}[Dynamic clustering criterion]
The set of clusters is re-estimated when at least one of the following conditions is true: a) the most recent window of samples is statistically different from the past windows, b) the most recent sample exceeds the range of previously observed data by a significant degree.
\end{definition}

The first case is a statistical deviation condition. For our anomaly detection method, this can be determined by the output of the detector itself. That is, the clusters are re-estimated every time an anomaly is detected. The correctness in this strategy is understood when we consider that, first, the early operation of the detector will declare most windows as anomalous until the detector learns enough patterns. At this stage, the clusters are highly inaccurate and need frequent updating. Second, both true and false positives are by definition unusual patterns and hence need to be considered by the discretization scheme. The second case is a spatial deviation condition. A natural scale of spatial deviation is the standard deviation $\sigma_K$, which is proportional to the smoothness parameter. For the Gaussian kernel, the standard deviation equals exactly the smoothness parameter, hence it can be defined by~\eqref{eq:Gauss_hopt}.

Overall, our proposed dynamic clustering method is based on a hybrid criterion for re-estimating the clusters, with means inherent to the density estimation and anomaly detection processes. The anomaly detector tracks the statistical deviation of the data stream, whilst the smoothness parameter of the implicit KDE process is used as a measure of spatial deviation. 


\section{Experimental Results}\label{sec:results}

\subsection{Experimental Evaluation of pSAX}\label{sec:psax_exp}

The quality of pSAX is evaluated and compared with its competitors in terms of the TLB~\eqref{eq:TLB} and RMSE~\eqref{eq:error} metrics. The evaluation is done on a selection of datasets, all of which are available in the UCR archive~\cite{bib:UCRArchive2018}. In particular, the \textit{Koski ECG} is an electrocardiogram dataset, the \textit{Muscle activation} is an electromyogram dataset, the \textit{Posture} is an object coordinates dataset and the \textit{Respiration} is a dataset of a patient's thorax extension. The criteria for choosing those datasets are that (i) they are not too small, as this would limit the variability between the subsequences, and (ii) they vary in structural content, such that they are representative for many other datasets.

In what follows, pSAX is compared with the conventional SAX and the adaptive SAX (aSAX)~\cite{bib:asax} methods. The latter is an alternative data-driven SAX method that employs the $k$-means algorithm for the discretization step. For the traditional SAX, which conventionally does not compute numerical values for the symbols, the RMSE~\eqref{eq:error} is computed by using the centroids of the equiprobable intervals under the Gaussian density function. The pSAX and aSAX are trained on a subset of the available PAA samples of size $\sqrt{L}$, where $L$ is the total number of the PAA samples. Note that, in order to obtain all of the available PAA samples, all subsequences of length $N$ must be transformed to their PAA representation.

The experimental evaluation is set up as follows. We randomly sample two subsequences from the same dataset $1000$ times for each combination of parameters. The subsequences are Z-normalized individually. Both subsequences are used to measure the TLB and one of them is used to measure the RMSE after its dimensionality reduction. The subsequence time series length $N$ varies in $\{480, 960, 1440, 1920\}$ and the number of bytes per subsequence varies in $\{8, 16, 24, 40\}$. The alphabet size $\kappa$ varies in $\{16, 256\}$, which is equivalent to 4 and 8 bits per symbol, respectively. Thus for each length $N$, the dimensionality $M$ of the symbolic representation varies in $\{16, 32, 48, 80\}$ for $\kappa = 16$ and in $\{8, 16, 24, 40\}$ for $\kappa = 256$. The results are displayed in Tables~\ref{tab:tlb1}-\ref{tab:rmse2}.

We observe that pSAX outperforms both SAX and aSAX in the vast majority of the experiments, in both TLB and RMSE scores. Also, we notice that the superiority of pSAX is more prominent when i) the number of available bytes is larger and ii) when the alphabet size is smaller. This behaviour is explained by the facts that a) a larger number of bytes provides more PAA samples for pre-training the KDE and Lloyd-Max quantizer, and b) an adequately large alphabet size drives any discretization method to produce very close discretization intervals. We emphasize that the speedup obtained by indexing with a higher TLB is higher than linear~\cite{bib:isax}. Thus we expect a greater speedup difference than the difference of TLB in Tables~\ref{tab:tlb1}-\ref{tab:tlb2}.

From Tables~\ref{tab:tlb1}-\ref{tab:rmse2}, one would infer that the choice of a SAX-based method is a matter of data source and/or parameterization. We assert that it is truly a matter of availability of training data. To this end, we verify this argument by comparing the lengths of the original datasets and the observation we made about the performance of our method versus the available bytes in the symbolic representation.

Interestingly, pSAX and aSAX should asymptotically converge to the same points, as the number of training data grows to infinity. This is because of the relationship between the Lloyd-Max quantizer and the $k$-means method (the latter is the finite version of the first), and because of the convergence property of KDE. However, in our experiments we illustrate the efficacy of the methods in the finite regime, and observe the generalization capability of kernel density estimators.

\begin{table*}[t]
\centering
\renewcommand{\arraystretch}{1.015}
\caption{Tightness of Lower Bound for Alphabet Size $\kappa = 256$}
\vspace{-10pt}
\label{tab:tlb1}
    \begin{tabular}{| c | c || c | c | c || c | c | c || c | c | c || c | c | c |}
        \hline
        & & \multicolumn{12}{c|}{\textbf{Available bytes}}\\
        \hline
        & & \multicolumn{3}{c||}{$8$} & \multicolumn{3}{c||}{$16$} & \multicolumn{3}{c||}{$24$} & \multicolumn{3}{c|}{$40$}\\
        \hline
        \textbf{Dataset} & \textbf{N} & pSAX & aSAX & SAX & pSAX & aSAX & SAX & pSAX & aSAX & SAX & pSAX & aSAX & SAX\\
        \hline
    \multirow{4}{*}{Koski ECG} & 480 & $0.560$ & ${0.560}$ & ${0.560}$ & $\mathbf{0.699}$ & $0.698$     & $0.697$ & $\mathbf{0.781}$ & $0.775$ & $0.769$ & $\mathbf{0.882}$ & $0.870$ & $0.828$\\
        & 960 & $\mathbf{0.385}$ & $0.382$ & $\mathbf{0.385}$ & $\mathbf{0.573}$ & $0.568$ & $\mathbf{0.573}$ & $\mathbf{0.652}$ & $0.646$ & $\mathbf{0.652}$ & $\mathbf{0.752}$ & $0.749$ & $0.749$\\
        & 1440 & $\mathbf{0.294}$ & $0.293$ & $0.293$ & ${0.481}$ & ${0.481}$ & ${0.481}$ & ${0.581}$ & ${0.581}$ & ${0.581}$ & ${0.678}$ & ${0.678}$ & ${0.678}$\\
        & 1920 & ${0.231}$ & ${0.231}$ & ${0.231}$ & ${0.403}$ & ${0.403}$ & ${0.403}$ & ${0.518}$ & ${0.518}$ & ${0.518}$ & ${0.626}$ & ${0.626}$ & ${0.626}$\\
        \hline
    \multirow{4}{*}{Muscle Activation} & 480 & $0.613$ & $\mathbf{0.615}$ & $\mathbf{0.615}$ & $0.710$ & $0.710$     & $0.710$ & $\mathbf{0.771}$ & $0.770$ & $0.770$ & $\mathbf{0.856}$ & $0.853$ & $0.853$\\
        & 960 & $0.776$ & $\mathbf{0.777}$ & $\mathbf{0.777}$ & $\mathbf{0.837}$ & $0.835$ & $0.836$ & $\mathbf{0.865}$ & $0.863$ & $0.863$ & $\mathbf{0.897}$ & $0.894$ & $0.893$\\
        & 1440 & $0.790$ & $0.790$ & $0.790$ & $\mathbf{0.851}$ & $0.850$ & $0.850$ & $\mathbf{0.873}$ & $0.872$ & $0.872$ & $\mathbf{0.900}$ & $0.899$ & $0.889$\\
        & 1920 & $\mathbf{0.763}$ & $0.762$ & $0.762$ & $\mathbf{0.837}$ & $0.835$ & $0.836$ & $\mathbf{0.863}$ & $0.862$ & $0.862$ & $\mathbf{0.890}$ & $0.889$ & $0.889$\\
        \hline
    \multirow{4}{*}{Posture} & 480 & $0.740$ & $0.740$ & $\mathbf{0.741}$ & $\mathbf{0.830}$ & $0.827$     & $0.827$ & $\mathbf{0.863}$ & $0.858$ & $0.857$ & $\mathbf{0.892}$ & $0.885$ & $0.880$\\
        & 960 & $0.665$ & $0.665$ & $\mathbf{0.666}$ & $\mathbf{0.786}$ & $0.782$ & $0.783$ & $\mathbf{0.835}$ & $0.830$ & $0.829$ & $\mathbf{0.874}$ & $0.867$ & $0.863$\\
        & 1440 & $0.616$ & $0.616$ & $\mathbf{0.617}$ & $\mathbf{0.751}$ & $0.750$ & $\mathbf{0.751}$ & $\mathbf{0.811}$ & $0.807$ & $0.807$ & $\mathbf{0.862}$ & $0.857$ & $0.855$\\
        & 1920 & $0.575$ & $0.576$ & $\mathbf{0.577}$ & $0.713$ & $0.713$ & $\mathbf{0.714}$ & $\mathbf{0.784}$ & $0.782$ & $0.782$ & $\mathbf{0.845}$ & $0.840$ & $0.838$\\
        \hline
    \multirow{4}{*}{Respiration} & 480 & $0.342$ & $\mathbf{0.344}$ & $\mathbf{0.344}$ & $0.464$ & $\mathbf{0.466}$     & $\mathbf{0.466}$ & $0.655$ & $\mathbf{0.656}$ & $0.655$ & $\mathbf{0.832}$ & $0.829$ & $0.827$\\
        & 960 & $0.281$ & $\mathbf{0.283}$ & $\mathbf{0.283}$ & $0.379$ & $\mathbf{0.381}$ & $\mathbf{0.381}$ & $0.389$ & $\mathbf{0.392}$ & $0.390$ & $\mathbf{0.594}$ & $\mathbf{0.594}$ & $0.590$\\
        & 1440 & $0.257$ & $\mathbf{0.259}$ & $\mathbf{0.259}$ & $0.336$ & $0.339$ & $\mathbf{0.340}$ & $0.399$ & $\mathbf{0.402}$ & $0.401$ & $\mathbf{0.439}$ & $0.438$ & $0.433$\\
        & 1920 & $0.239$ & $0.240$ & $\mathbf{0.241}$ & $0.324$ & $\mathbf{0.327}$ & $\mathbf{0.327}$ & $0.366$ & $\mathbf{0.369}$ & $\mathbf{0.369}$ & $\mathbf{0.422}$ & $0.419$ & $0.415$\\
        \hline
    \end{tabular}
    \vspace{-4pt}
\end{table*}

\begin{table*}[t]
\centering
\renewcommand{\arraystretch}{1.015}
\caption{Tightness of Lower Bound for Alphabet Size $\kappa = 16$}
\vspace{-10pt}
\label{tab:tlb2}
    \begin{tabular}{| c | c || c | c | c || c | c | c || c | c | c || c | c | c |}
        \hline
        & & \multicolumn{12}{c|}{\textbf{Available bytes}}\\
        \hline
        & & \multicolumn{3}{c||}{$8$} & \multicolumn{3}{c||}{$16$} & \multicolumn{3}{c||}{$24$} & \multicolumn{3}{c|}{$40$}\\
        \hline
        \textbf{Dataset} & \textbf{N} & pSAX & aSAX & SAX & pSAX & aSAX & SAX & pSAX & aSAX & SAX & pSAX & aSAX & SAX\\
        \hline
    \multirow{4}{*}{Koski ECG} & 480 & $\mathbf{0.641}$ & $0.638$ & $0.629$ & $0.774$ & $\mathbf{0.775}$     & $0.707$ & $\mathbf{0.835}$ & $\mathbf{0.835}$ & $0.747$ & $\mathbf{0.883}$ & $\mathbf{0.883}$ & $0.778$\\
        & 960 & $\mathbf{0.531}$ & $0.529$ & $0.523$ & $\mathbf{0.656}$ & $0.652$ & $0.642$ & $\mathbf{0.727}$ & $0.721$ & $0.682$ & $\mathbf{0.819}$ & $0.817$ & $0.737$\\
        & 1440 & $\mathbf{0.441}$ & $0.437$ & $0.433$ & $\mathbf{0.588}$ & $0.587$ & $0.581$ & $\mathbf{0.658}$ & $0.655$ & $0.645$ & $\mathbf{0.753}$ & $\mathbf{0.753}$ & $0.698$\\
        & 1920 & $\mathbf{0.369}$ & $0.367$ & $0.360$ & $\mathbf{0.540}$ & $0.539$ & $0.535$ & $\mathbf{0.613}$ & $0.611$ & $0.605$ & $\mathbf{0.701}$ & $0.697$ & $0.670$\\
        \hline
    \multirow{4}{*}{Muscle Activation} & 480 & $\mathbf{0.637}$ & $0.632$ & $\mathbf{0.637}$ & $\mathbf{0.744}$ & $0.737$     & $0.742$ & $\mathbf{0.798}$ & $0.790$ & $0.794$ & $\mathbf{0.848}$ & $0.839$ & $0.841$\\
        & 960 & $0.756$ & $\mathbf{0.758}$ & $0.751$ & $\mathbf{0.799}$ & $0.798$ & $0.790$ & $\mathbf{0.823}$ & $\mathbf{0.823}$ & $0.814$ & $\mathbf{0.852}$ & $\mathbf{0.852}$ & $0.842$\\
        & 1440 & $0.798$ & $\mathbf{0.799}$ & $0.783$ & $0.835$ & $\mathbf{0.836}$ & $0.821$ & $\mathbf{0.823}$ & $\mathbf{0.823}$ & $0.814$ & $\mathbf{0.852}$ & $\mathbf{0.852}$ & $0.842$\\
        & 1920 & $0.789$ & $\mathbf{0.791}$ & $0.771$ & $0.833$ & $\mathbf{0.834}$ & $0.814$ & $0.850$ & $\mathbf{0.851}$ & $0.831$ & $\mathbf{0.870}$ & $\mathbf{0.870}$ & $0.852$\\
        \hline
    \multirow{4}{*}{Posture} & 480 & $\mathbf{0.751}$ & $0.744$ & $0.736$ & $\mathbf{0.798}$ & $0.787$ & $0.774$ & $\mathbf{0.815}$ & $0.807$ & $0.787$ & $\mathbf{0.835}$ & $0.825$ & $0.803$\\
        & 960 & $\mathbf{0.716}$ & $0.710$ & $0.703$ & $\mathbf{0.782}$ & $0.781$ & $0.753$ & $\mathbf{0.810}$ & $0.805$ & $0.773$ & $\mathbf{0.833}$ & $0.830$ & $0.790$\\
        & 1440 & $\mathbf{0.685}$ & $0.681$ & $0.677$ & $\mathbf{0.768}$ & $0.765$ & $0.742$ & $\mathbf{0.801}$ & $0.798$ & $0.767$ & $\mathbf{0.830}$ & $0.827$ & $0.788$\\
        & 1920 & $\mathbf{0.650}$ & $0.650$ & $0.647$ & $\mathbf{0.748}$ & $0.747$ & $0.728$ & $\mathbf{0.787}$ & $0.786$ & $0.757$ & $\mathbf{0.820}$ & $0.819$ & $0.781$\\
        \hline
    \multirow{4}{*}{Respiration} & 480 & $\mathbf{0.420}$ & $0.417$ & $0.415$ & $\mathbf{0.705}$ & $0.695$     & $0.703$ & $\mathbf{0.805}$ & $0.801$ & $0.784$ & $\mathbf{0.872}$ & $0.869$ & $0.838$\\
        & 960 & $\mathbf{0.340}$ & $0.335$ & $0.334$ & $\mathbf{0.442}$ & $0.441$ & $0.436$ & $\mathbf{0.612}$ & $0.605$ & $0.603$ & $\mathbf{0.766}$ & $0.765$ & $0.748$\\
        & 1440 & $\mathbf{0.300}$ & $0.297$ & $0.295$ & $\mathbf{0.359}$ & $0.354$ & $0.350$ & $0.458$ & $\mathbf{0.460}$ & $0.449$ & $\mathbf{0.658}$ & $0.652$ & $0.644$\\
        & 1920 & $\mathbf{0.291}$ & $0.289$ & $0.285$ & $\mathbf{0.374}$ & $0.370$ & $0.362$ & $\mathbf{0.386}$ & $0.379$ & $0.373$ & $0.558$ & $\mathbf{0.559}$ & $0.545$\\
        \hline
    \end{tabular}
    \vspace{-1pt}
\end{table*}

\begin{table*}[t]
\centering
\renewcommand{\arraystretch}{1.015}
\caption{Root Mean Squared Error for Alphabet Size $\kappa = 256$}
\vspace{-10pt}
\label{tab:rmse1}
    \begin{tabular}{| c | c || c | c | c || c | c | c || c | c | c || c | c | c |}
        \hline
        & & \multicolumn{12}{c|}{\textbf{Available bytes}}\\
        \hline
        & & \multicolumn{3}{c||}{$8$} & \multicolumn{3}{c||}{$16$} & \multicolumn{3}{c||}{$24$} & \multicolumn{3}{c|}{$40$}\\
        \hline
        \textbf{Dataset} & \textbf{N} & pSAX & aSAX & SAX & pSAX & aSAX & SAX & pSAX & aSAX & SAX & pSAX & aSAX & SAX\\
        \hline
    \multirow{4}{*}{Koski ECG} & 480 & ${0.800}$ & ${0.800}$ & ${0.800}$ & ${0.685}$ & ${0.685}$     & ${0.685}$ & $\mathbf{0.587}$ & $\mathbf{0.587}$ & $0.588$ & $\mathbf{0.433}$ & $0.434$ & $0.464$\\
        & 960 & ${0.903}$ & ${0.903}$ & ${0.903}$ & ${0.799}$ & ${0.799}$ & ${0.799}$ & ${0.737}$ & ${0.737}$ & ${0.737}$ & ${0.636}$ & ${0.636}$ & ${0.636}$\\
        & 1440 & ${0.944}$ & ${0.944}$ & ${0.944}$ & ${0.859}$ & ${0.859}$ & ${0.859}$ & ${0.796}$ & ${0.796}$ & ${0.796}$ & ${0.717}$ & ${0.717}$ & ${0.717}$\\
        & 1920 & ${0.963}$ & ${0.963}$ & ${0.963}$ & ${0.899}$ & ${0.899}$ & ${0.899}$ & ${0.839}$ & ${0.839}$ & ${0.839}$ & ${0.761}$ & ${0.761}$ & ${0.761}$\\
        \hline
    \multirow{4}{*}{Muscle Activation} & 480 & $0.694$ & $0.694$ & $0.694$ & $0.611$ & $0.611$     & $0.611$ & $0.644$ & $0.644$ & $0.644$ & $\mathbf{0.435}$ & $0.436$ & $0.436$\\
        & 960 & $0.504$ & $0.504$ & $0.504$ & $0.533$ & $0.533$ & $0.533$ & $0.394$ & $0.394$ & $0.394$ & $\mathbf{0.343}$ & $0.344$ & $0.344$\\
        & 1440 & $0.484$ & $0.484$ & $0.484$ & $0.406$ & $0.406$ & $0.406$ & $0.372$ & $0.372$ & $0.372$ & $0.331$ & $0.331$ & $0.331$\\
        & 1920 & $0.528$ & $0.528$ & $0.528$ & $0.430$ & $0.430$ & $0.430$ & $0.391$ & $0.391$ & $0.391$ & $0.351$ & $0.351$ & $0.351$\\
        \hline
    \multirow{4}{*}{Posture} & 480 & $0.577$ & $0.577$ & $0.577$ & $0.454$ & $0.454$ & $0.454$ & $\mathbf{0.396}$ & $0.398$ & $0.398$ & $\mathbf{0.336}$ & $0.338$ & $0.342$\\
        & 960 & $0.693$ & $0.693$ & $0.693$ & $\mathbf{0.560}$ & $0.561$ & $\mathbf{0.560}$ & $\mathbf{0.487}$ & $0.488$ & $0.488$ & $\mathbf{0.418}$ & $0.419$ & $0.420$\\
        & 1440 & $0.751$ & $0.751$ & $0.751$ & $0.618$ & $0.618$ & $0.618$ & $\mathbf{0.538}$ & $0.539$ & $0.539$ & $\mathbf{0.461}$ & $\mathbf{0.461}$ & $0.462$\\
        & 1920 & $0.785$ & $0.785$ & $0.785$ & $0.663$ & $0.663$     & $0.663$ & $0.584$ & $0.584$ & $0.584$ & $\mathbf{0.499}$ & $\mathbf{0.499}$ & $0.500$\\
        \hline
    \multirow{4}{*}{Respiration} & 480 & $0.930$ & $0.930$ & $0.930$ & $0.874$ & $\mathbf{0.873}$     & $0.874$ & $\mathbf{0.735}$ & $0.736$ & $0.736$ & $0.534$ & $0.535$ & $0.537$\\
        & 960 & $0.951$ & $0.951$ & $0.951$ & $0.915$ & $0.915$ & $0.915$ & $\mathbf{0.906}$ & $\mathbf{0.906}$ & $0.907$ & $\mathbf{0.787}$ & $\mathbf{0.787}$ & $0.791$\\
        & 1440 & $0.959$ & $0.959$ & $0.959$ & $0.932$ & $0.932$ & $0.932$ & $0.904$ & $0.904$ & $0.904$ & $\mathbf{0.880}$ & $\mathbf{0.880}$ & $0.884$\\
        & 1920 & $0.964$ & $0.964$ & $0.964$ & $0.937$ & $0.937$ & $0.937$ & $0.919$ & $\mathbf{0.918}$ & $\mathbf{0.918}$ & $\mathbf{0.890}$ & $0.891$ & $0.894$\\
        \hline
    \end{tabular}
    \vspace{-4pt}
\end{table*}

\begin{table*}[t]
\centering
\renewcommand{\arraystretch}{1.015}
\caption{Root Mean Squared Error for Alphabet Size $\kappa = 16$}
\vspace{-10pt}
\label{tab:rmse2}
    \begin{tabular}{| c | c || c | c | c || c | c | c || c | c | c || c | c | c |}
        \hline
        & & \multicolumn{12}{c|}{\textbf{Available bytes}}\\
        \hline
        & & \multicolumn{3}{c||}{$8$} & \multicolumn{3}{c||}{$16$} & \multicolumn{3}{c||}{$24$} & \multicolumn{3}{c|}{$40$}\\
        \hline
        \textbf{Dataset} & \textbf{N} & pSAX & aSAX & SAX & pSAX & aSAX & SAX & pSAX & aSAX & SAX & pSAX & aSAX & SAX\\
        \hline
    \multirow{4}{*}{Koski ECG} & 480 & $\mathbf{0.688}$ & $\mathbf{0.688}$ & $0.690$ & $\mathbf{0.506}$ & $\mathbf{0.506}$     & $0.569$ & $\mathbf{0.388}$ & $\mathbf{0.388}$ & $0.518$ & $\mathbf{0.265}$ & $\mathbf{0.265}$ & $0.479$\\
        & 960 & $\mathbf{0.800}$ & $0.801$ & $0.801$ & $\mathbf{0.688}$ & $\mathbf{0.688}$ & $0.690$ & $\mathbf{0.594}$ & $0.595$ & $0.613$ & $\mathbf{0.445}$ & $\mathbf{0.445}$ & $0.538$\\
        & 1440 & $0.863$ & $0.863$ & $0.863$ & $\mathbf{0.757}$ & $\mathbf{0.757}$ & $0.758$ & $\mathbf{0.687}$ & $\mathbf{0.687}$ & $0.689$ & $\mathbf{0.565}$ & $\mathbf{0.565}$ & $0.595$\\
        & 1920 & $0.900$ & $0.900$ & $0.900$ & $\mathbf{0.796}$ & $0.797$ & $0.797$ & $\mathbf{0.736}$ & $\mathbf{0.736}$ & $0.738$ & $\mathbf{0.638}$ & $\mathbf{0.638}$ & $0.644$\\
        \hline
    \multirow{4}{*}{Muscle Activation} & 480 & $\mathbf{0.616}$ & $0.617$ & $0.619$ & $\mathbf{0.496}$ & $0.498$     & $0.503$ & $\mathbf{0.413}$ & $0.417$ & $0.425$ & $\mathbf{0.309}$ & $0.313$ & $0.326$\\
        & 960 & $\mathbf{0.441}$ & $\mathbf{0.441}$ & $0.447$ & $\mathbf{0.378}$ & $\mathbf{0.378}$ & $0.389$ & $\mathbf{0.334}$ & $0.335$ & $0.347$ & $\mathbf{0.273}$ & $\mathbf{0.273}$ & $0.292$\\
        & 1440 & $\mathbf{0.409}$ & $\mathbf{0.409}$ & $0.412$ & $\mathbf{0.355}$ & $\mathbf{0.355}$ & $0.359$ & $\mathbf{0.322}$ & $\mathbf{0.322}$ & $0.328$ & $\mathbf{0.273}$ & $\mathbf{0.273}$ & $0.281$\\
        & 1920 & $\mathbf{0.432}$ & $\mathbf{0.432}$ & $0.435$ & $\mathbf{0.372}$ & $\mathbf{0.372}$ & $0.376$ & $\mathbf{0.341}$ & $\mathbf{0.341}$ & $0.345$ & $\mathbf{0.298}$ & $\mathbf{0.298}$ & $0.304$\\
        \hline
    \multirow{4}{*}{Posture} & 480 & $\mathbf{0.469}$ & $0.472$ & $0.480$ & $\mathbf{0.376}$ & $0.379$     & $0.410$ & $\mathbf{0.342}$ & $0.347$ & $0.378$ & $\mathbf{0.290}$ & $0.297$ & $0.342$\\
        & 960 & $\mathbf{0.562}$ & $0.564$ & $0.570$ & $\mathbf{0.457}$ & $0.459$ & $0.478$ & $\mathbf{0.407}$ & $0.408$ & $0.436$ & $\mathbf{0.354}$ & $0.356$ & $0.396$\\
        & 1440 & $\mathbf{0.621}$ & $0.622$ & $0.624$ & $\mathbf{0.504}$ & $0.505$ & $0.517$ & $\mathbf{0.407}$ & $0.408$ & $0.436$ & $\mathbf{0.354}$ & $0.356$ & $0.396$\\
        & 1920 & $\mathbf{0.667}$ & $\mathbf{0.667}$ & $0.669$ & $\mathbf{0.541}$ & $\mathbf{0.541}$ & $0.550$ & $\mathbf{0.478}$ & $\mathbf{0.478}$ & $0.493$ & $\mathbf{0.414}$ & $\mathbf{0.414}$ & $0.438$\\
        \hline
    \multirow{4}{*}{Respiration} & 480 & $\mathbf{0.877}$ & $0.880$ & $0.878$ & $\mathbf{0.627}$ & $0.629$     & $0.635$ & $\mathbf{0.476}$ & $0.478$ & $0.493$ & $\mathbf{0.319}$ & $0.323$ & $0.356$\\
        & 960 & $\mathbf{0.917}$ & $\mathbf{0.917}$ & $0.918$ & $\mathbf{0.860}$ & $0.866$ & $0.866$ & $\mathbf{0.725}$ & $\mathbf{0.725}$ & $0.738$ & $\mathbf{0.530}$ & $0.533$ & $0.559$\\
        & 1440 & $0.934$ & $0.934$ & $0.934$ & $\mathbf{0.903}$ & $0.905$ & $0.910$ & $\mathbf{0.845}$ & $0.851$ & $0.859$ & $\mathbf{0.675}$ & $0.677$ & $0.702$\\
        & 1920 & $\mathbf{0.938}$ & $\mathbf{0.938}$ & $0.939$ & $\mathbf{0.898}$ & $\mathbf{0.898}$ & $0.904$ & $\mathbf{0.886}$ & $0.888$ & $0.899$ & $\mathbf{0.768}$ & $0.775$ & $0.795$\\
        \hline
    \end{tabular}
    \vspace{-1pt}
\end{table*}

\subsection{Experimental Evaluation of cSAX}

As already mentioned, the performance of cSAX is evaluated indirectly on the task of anomaly detection. To this end, we utilize the anomaly detector described by Algorithm~\ref{alg:anomaly}, hereafter referred to as the goodness of fit (GoF) detector, and the discord discovery algorithm HOT-SAX (ref. Sec.~\ref{sec:HOTSAX}). The choice of the GoF detector is based on the idea that, due to its theoretical and algorithmic simplicity, it will highlight the differences among the distinct time series representations. On the other hand, the HOT-SAX algorithm is the most widely used method for exact discord discovery and its speed depends highly on the quality of the SAX representation.

The GoF detector depends on two parameters; the alphabet size $\kappa$ and the window length $n$. In our experiments, the window length is set empirically to $n=50$, fixed for all datasets. Accordingly, the alphabet size is set to $\kappa=10$, except for cSAX that automatically detects the number of clusters. The HOT-SAX algorithm requires setting the discord length $l$, which is implied by the specific application or the data domain, and also requires careful setting of the SAX parameters, since it benefits from very low alphabet size $\kappa$ and dimensionality $M$. In our experiments, we set empirically the best parameters as $M=3$ and $\kappa = 3$, for all datasets. The alphabet size $\kappa$ cannot be fixed for the cSAX representation, but we can impose a low alphabet size by increasing the smoothness parameter in~\eqref{eq:Gauss_grad_hopt}. We did this by increasing the AMISE optimal setting~\cite{bib:HOTSAX} by a multiplication factor of 4. This resulted in alphabet sizes between 2 and 4 for most datasets, whilst for a limited number of them, up to the range of 10-15.

\subsubsection{Anomaly Detection with cSAX}

The anomaly detector is evaluated via receiver operating characteristic (ROC) curves and the corresponding area under the curve (AUC) values. The ROC curves and the AUC illustrate the efficacy of the detector to discriminate correctly between anomalous and normal samples. The datasets we employ belong to the Numenta Anomaly Benchmark (NAB)~\cite{bib:NAB}, which consists of 58 real-world and synthetic labelled datasets for anomaly detection. NAB also simplifies the comparison of our detector with other top-tier online anomaly detectors that are already implemented in NAB's library. 

In the subsequent experiments, some training sets are utilized. Notice though that, since we perform unsupervised anomaly detection, the training process does not refer to the detector, but to the underlying symbolic representation. In particular, SAX needs to learn the mean value and the standard deviation of the time series, pSAX learns the density function and the discretization intervals, and aSAX requires to learn the discretization intervals. The dynamic cSAX does not require training but allows pre-training for better initialization. To create the training sets, we provide a portion from the first section of the datasets as a training set. Then, the anomaly detection process is performed on the whole dataset.

Note that many detectors available in NAB's library make use of the minimum and maximum values of each dataset. That is also true for the detector we utilize in our experiments, due to the uniform quantization step. The knowledge of this information violates the principle of online (streaming) algorithms. The only argument in favour of this choice is that the minimum and maximum operating values are often known \`{a} priori. However, apart from the fact that this is not always true, it is not guaranteed that the datasets' minimum and maximum values coincide with the operating range, and the first provides more significant information than the latter. Our anomaly detector is capable of operating in a truly online fashion, thanks to the dynamic mean-shift clustering described in Sec.~\ref{sec:dynamic_MS}. At the same time, it is capable of pre-training the clustering on historical data, if available.

\begin{figure}
    \centering
    \includegraphics[width=.99\linewidth]{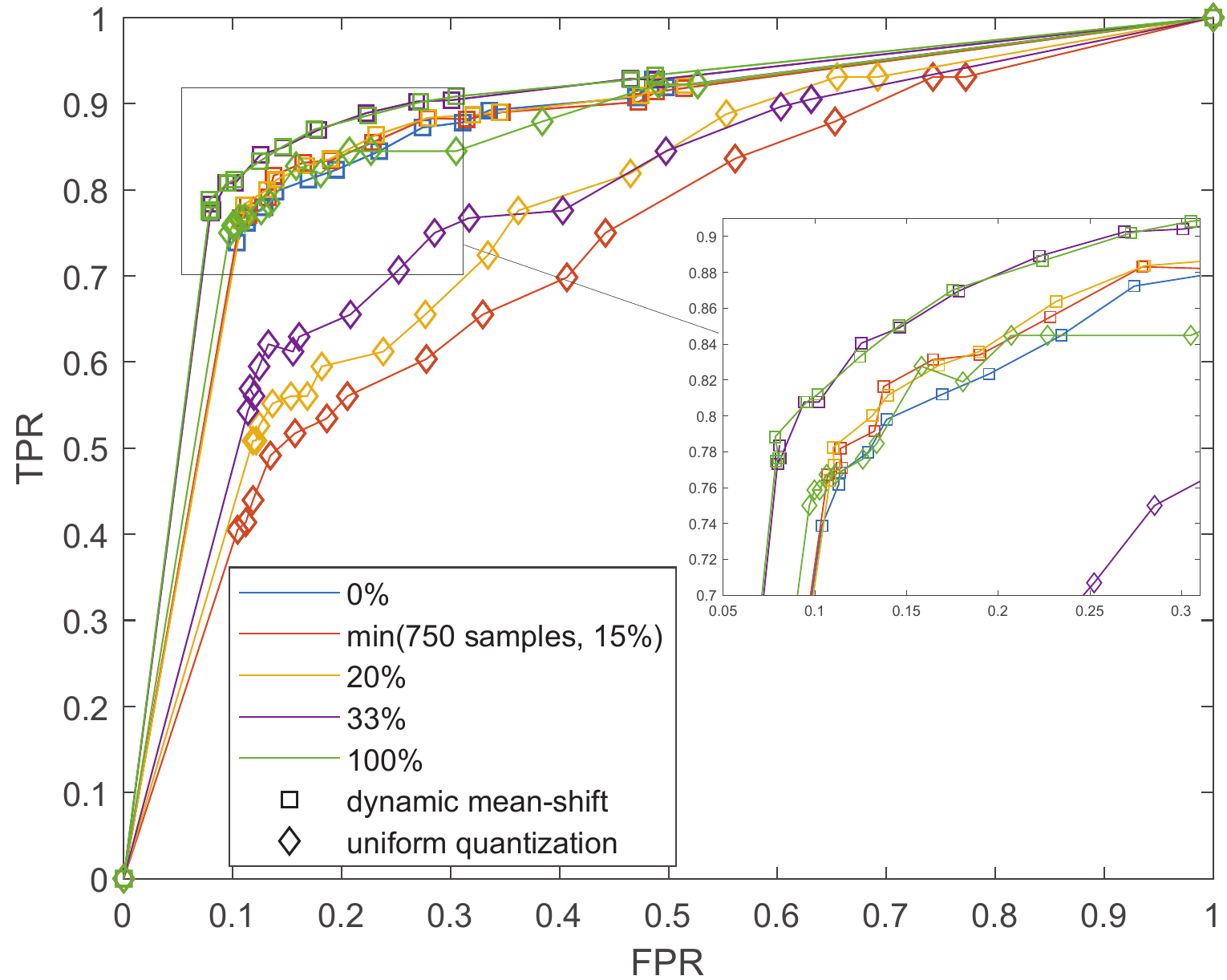}
    \caption{ROC curves for different training set sizes, with no dimensionality reduction. The training set sizes are expressed in percentages of the original dataset. Note that for the uniform quantizer, the optimal average alphabet size is set empirically ($\kappa = 10$, fixed for all datasets), whilst the mean-shift automatically detects the number of clusters.}
    \label{fig:ROC_vs_training}
\end{figure}

In the following, we consider the effect of different training set sizes on the detector's performance. The ROC curves and AUC values reported below are the weighted averages across all datasets, weighted by their length.  Fig.~\ref{fig:ROC_vs_training} compares the conventional GoF detector, with uniform quantization, and the GoF detector paired with the cSAX, with dynamic mean-shift clustering. For the cSAX we set $M=N$, that is, without dimensionality reduction. We emphasize again that the training set is used to estimate the clusters for cSAX and the minimum/maximum values for the uniform quantizer, but not the detector's model.

It can be seen that the dynamic mean-shift achieves much higher performance for small training set sizes, whilst the performance of the uniform quantizer with $100\%$ training data is practically equal to the performance of the dynamic mean-shift with $0\%$ training data. Importantly, the alphabet size is automatically detected for the dynamic mean-shift method, whilst an empirical search is required to find the best alphabet size for the uniform quantizer. Another important observation is that our dynamic clustering criterion is capable of achieving very similar performance between the $0\%$ and $100\%$ training set sizes.

Table~\ref{tab:nab} shows the NAB scores of our method against the top-performing methods from NAB's library. NAB calculates three different scores by weighing false positives (FPs) and false negatives (FNs) differently: (i) favouring fewer FPs, (ii) favouring fewer FNs, and (iii) balancing both FPs and FNs. The scores' values vary between $0$ and $100$ (the higher the better). The methods that require knowledge of the minimum and maximum values are noted with an asterisk. We evaluate cSAX with no pre-training, for a truly online application scenario. Note that the Numenta HTM exploits the minimum and maximum values, but can learn them online, so it is evaluated in a fully online fashion. Notably, the fully online Numenta HTM demonstrates a large decrease in scores, compared to the ones obtained conventionally\footnote{Available at: \url{https://github.com/numenta/NAB#scoreboard}}. Again, we observe that cSAX with no prior information and no need to tune the alphabet size, achieves the same performance as the conventional GoF detector with access to the true minimum and maximum values of the data and having empirically set the best alphabet size.

\begin{table}
\centering
\renewcommand{\arraystretch}{1.015}
\begin{threeparttable}
\caption{NAB Scores of the cSAX-GoF Detector Without Dimensionality Reduction Against its Competitors \vspace{-4pt}}
\label{tab:nab}
    \begin{tabular}{| c | c | c | c |}
        \hline
        \textbf{Detector} & \textbf{Standard} & \textbf{Low FP} & \textbf{Low FN}\\
        \hline
        {CAD OSE\tnote{*}} & $69.9$ & $67.0$ & $73.2$\\
        \hline
        {GoF\tnote{*}} & $60.4$ & $50.3$ & $67.6$\\
        \hline
        \textbf{cSAX-GoF} & $60.4$ & $49.6$ & $66.0$\\
        \hline
        {Numenta HTM} & $60.0$ & $53.8$ & $63.8$\\
        \hline
        {earthgecko Skyline} & $58.2$ & $46.2$ & $63.9$\\
        \hline
        {KNN CAD} & $58.0$ & $43.4$ & $64.8$\\
        \hline
        {Random Cut Forest} & $51.7$ & $38.4$ & $59.7$\\
        \hline
    \end{tabular}
\smallskip
\scriptsize
\begin{tablenotes}
\item[*] Require the minimum and maximum values of the data.
\end{tablenotes}
\end{threeparttable}
\vspace{-5pt}
\end{table}

\begin{figure}[t]
    \centering
    \subfloat[]{\includegraphics[width=.5\linewidth]{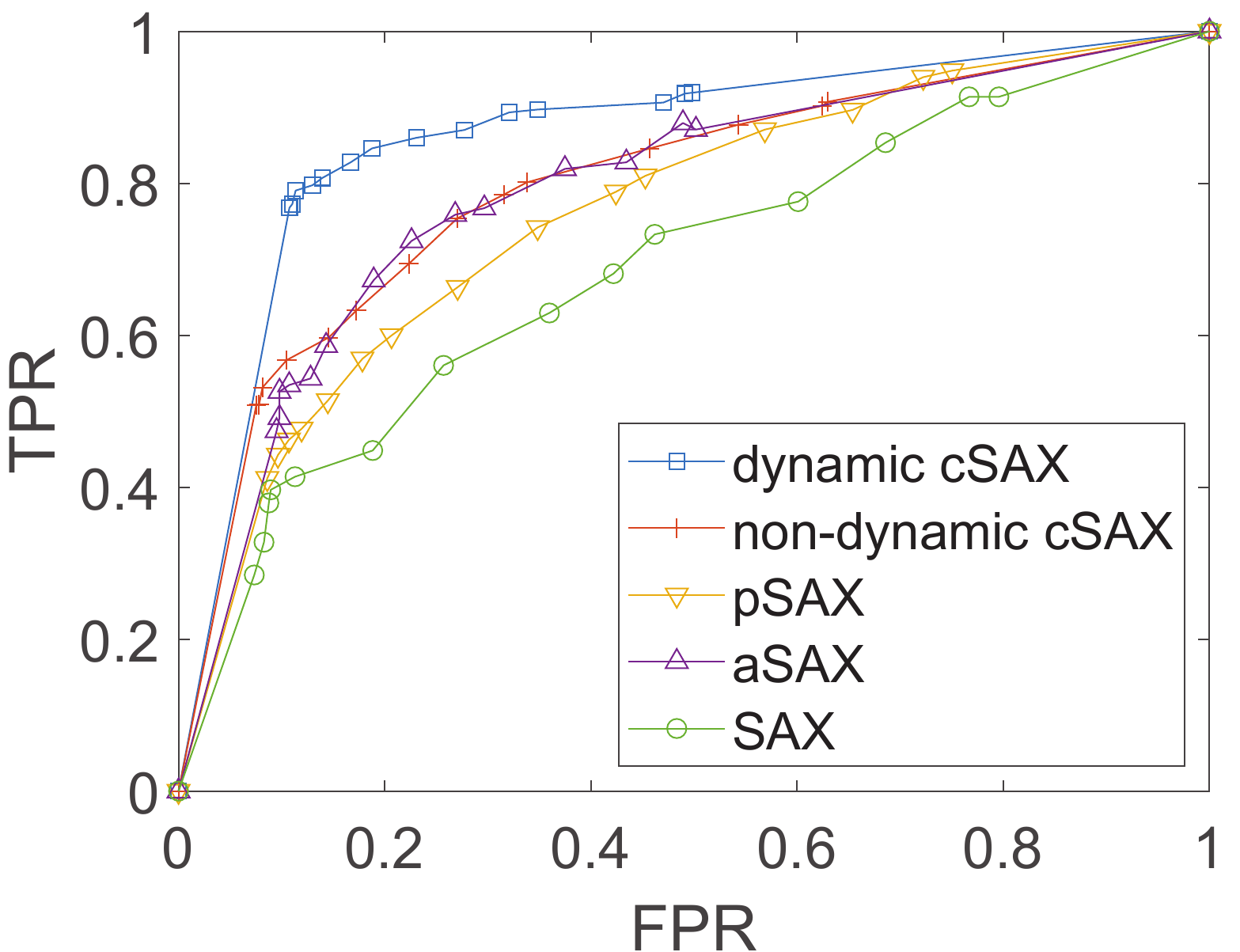}%
    }
    \hfil
    \subfloat[]{\includegraphics[width=.5\linewidth]{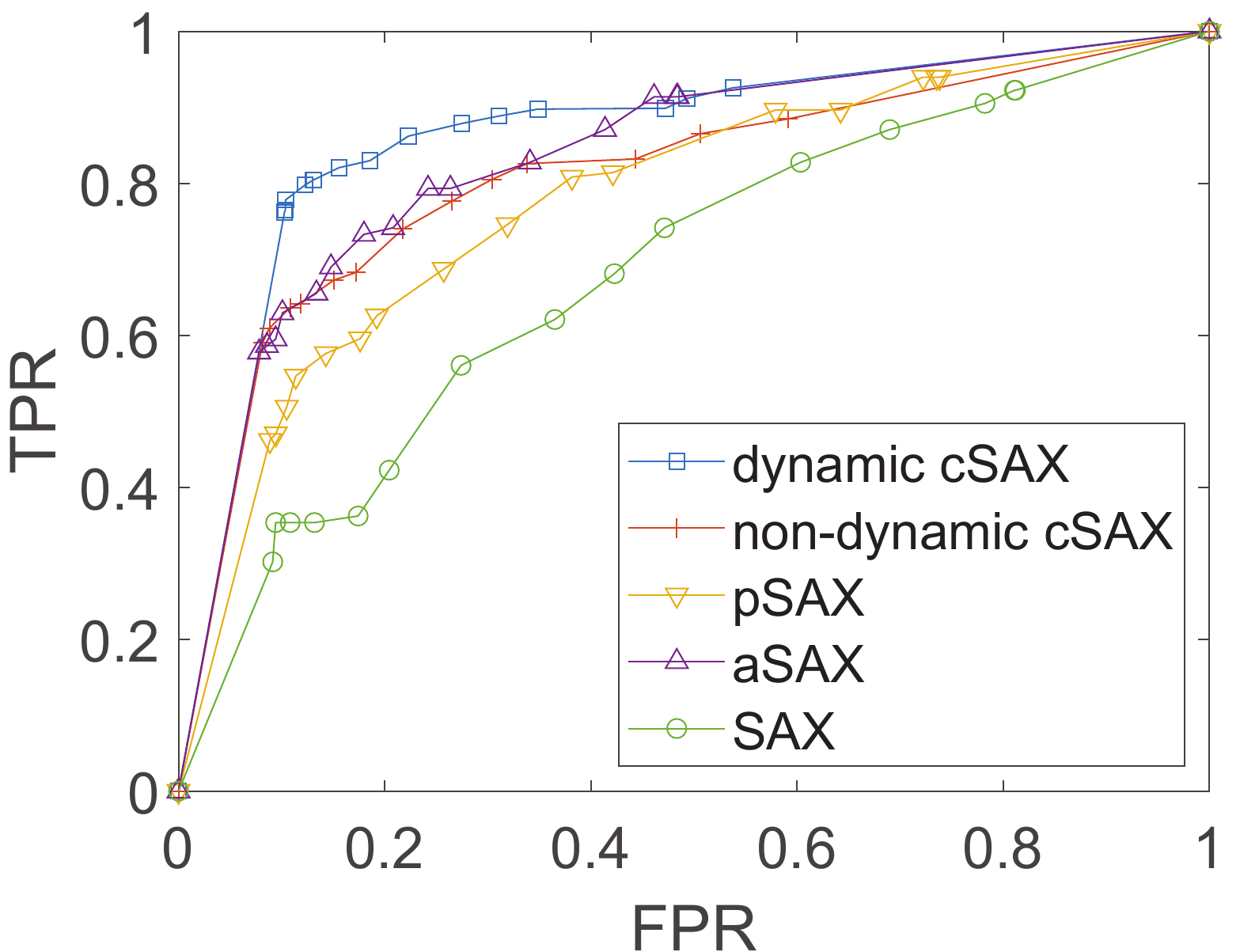}%
    }\\ \vspace{-7pt}
    \subfloat[]{\includegraphics[width=.5\linewidth]{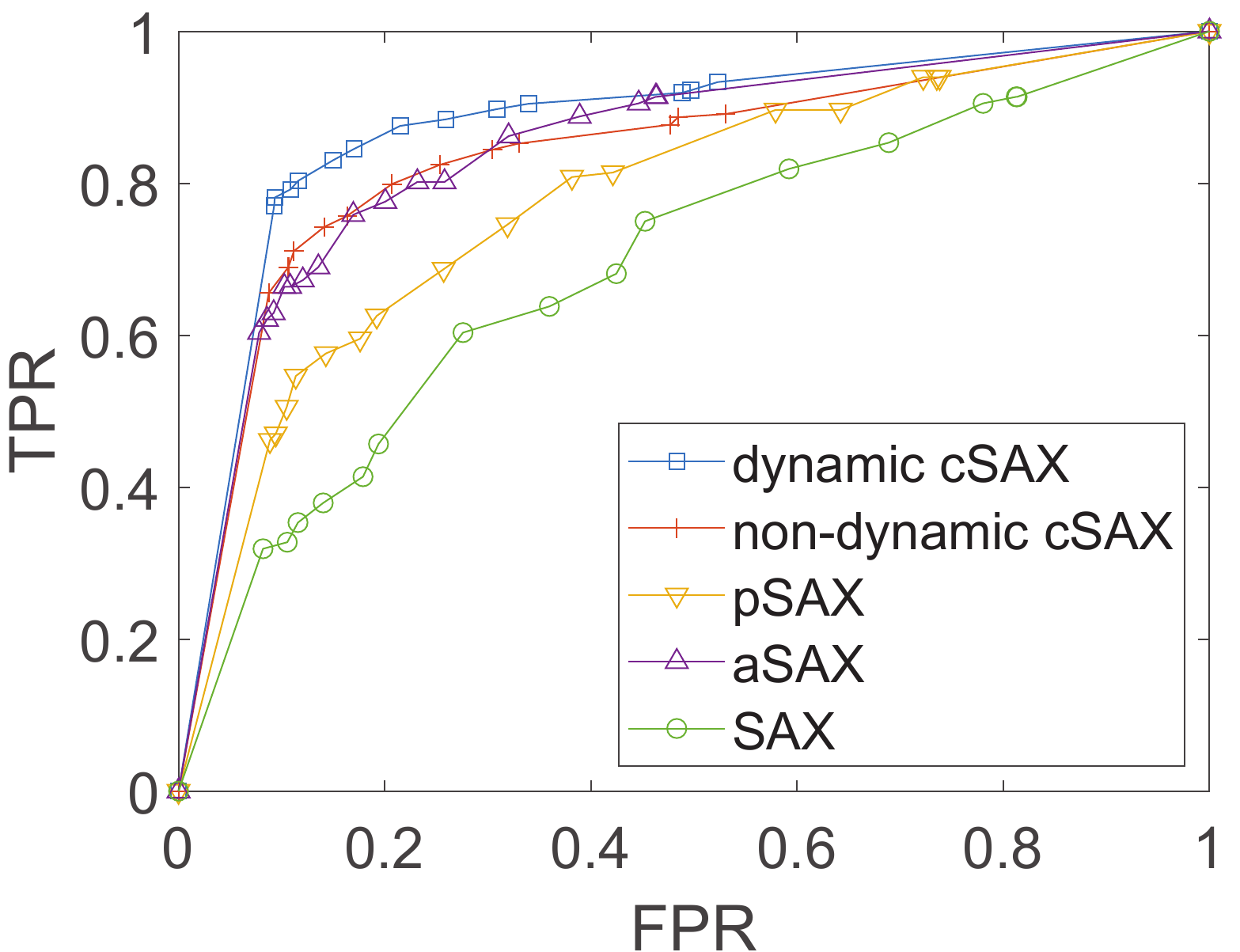}%
    }
    \hfil
    \subfloat[]{\includegraphics[width=.5\linewidth]{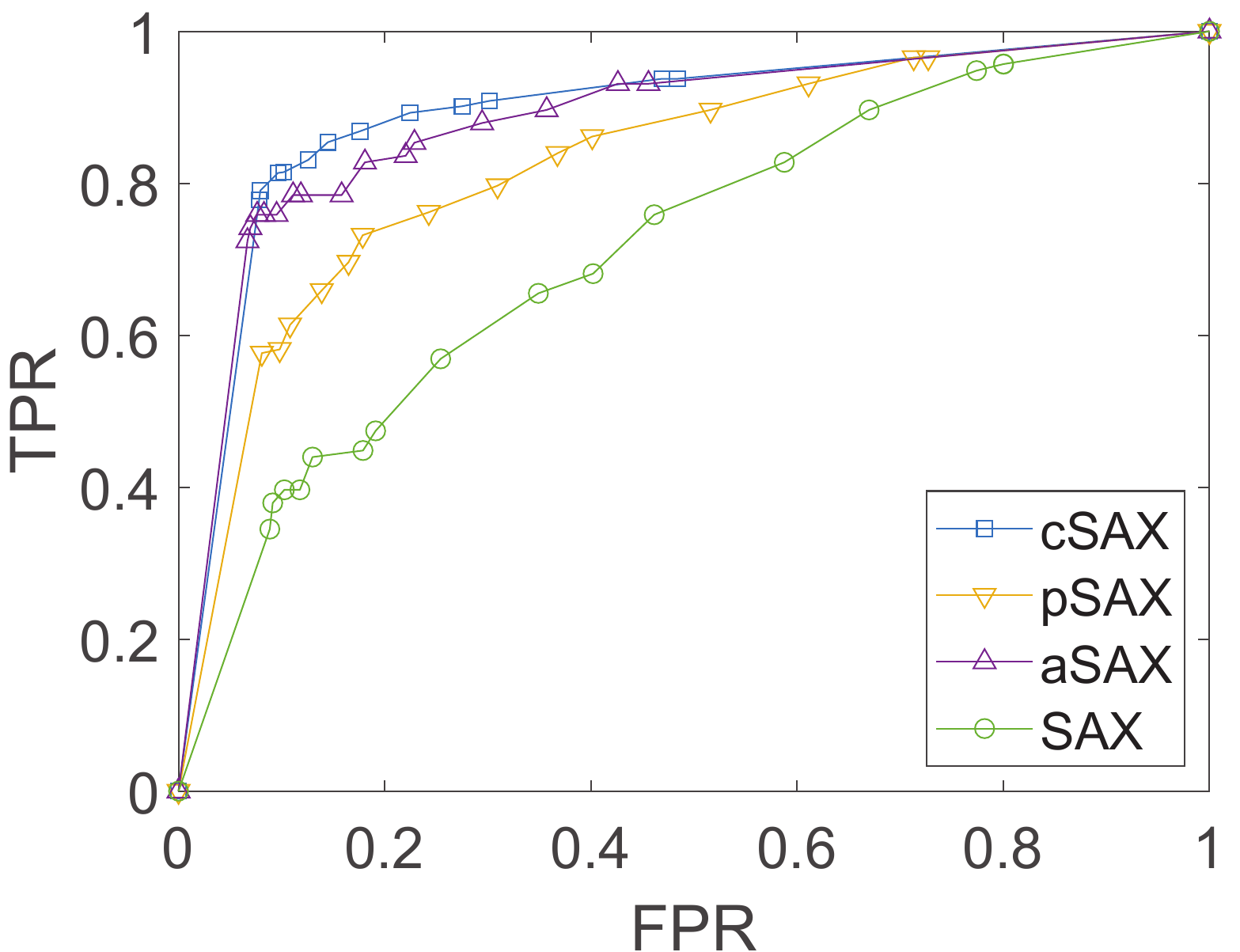}%
    }\\
    \smallskip\smallskip
    \begin{tabular}{| l | c | c | c | c | c |}
    \hline
    \footnotesize{Training set} & \footnotesize{dyn. cSAX} & \footnotesize{cSAX} & \footnotesize{pSAX} & \footnotesize{aSAX} & \footnotesize{SAX}\\
    \hline
    \footnotesize{(a) $20\%$} & \footnotesize{$\mathbf{0.8653}$} & \footnotesize{0.7996} & \footnotesize{0.7624} & \footnotesize{0.7986} & \footnotesize{0.7045}\\
    \hline
    \footnotesize{(b) $33\%$} & \footnotesize{$\mathbf{0.8673}$} & \footnotesize{0.8147} & \footnotesize{0.7840} & \footnotesize{0.8401} & \footnotesize{0.6949}\\
    \hline
    \footnotesize{(c) $66\%$} & \footnotesize{$\mathbf{0.8787}$} & \footnotesize{0.8406} & \footnotesize{0.7839} & \footnotesize{0.8495} & \footnotesize{0.7031}\\
    \hline
    \footnotesize{(d) $100\%$} & \multicolumn{2}{c|}{\footnotesize{$\mathbf{0.8934}$}} & \footnotesize{0.8320} & \footnotesize{0.8848} & \footnotesize{0.7224}\\
    \hline
    \end{tabular}
    \caption{ROC curves (top) and AUC values (bottom table) of the SAX-based representations (without dimensionality reduction) on NAB's datasets for different training set sizes: a) 20\% of total samples, b) 33\% of total samples, c) 66\% of total samples, d) 100\% of total samples.
    \label{fig:symbolic_ROC_no_dim}}
\end{figure}

Our goal at this point is not to achieve the best performance, but to compare how the underlying representation affects the performance of the same detector, which is the reason we chose a rather simple detector for our experiments. In Fig.~\ref{fig:symbolic_ROC_no_dim}, we compare the ROC curves for cSAX, pSAX, aSAX and SAX, without dimensionality reduction, for a varying training set size. We observe that the conventional SAX has the worst performance in all cases. The aSAX and the non-dynamic cSAX perform equally well, both better than pSAX and SAX. But only cSAX is able to detect automatically the alphabet size. More importantly, the superiority of the dynamic cSAX against all the other representations is apparent.

\begin{figure}[t]
    \centering
    \subfloat[]{\includegraphics[width=.5\linewidth]{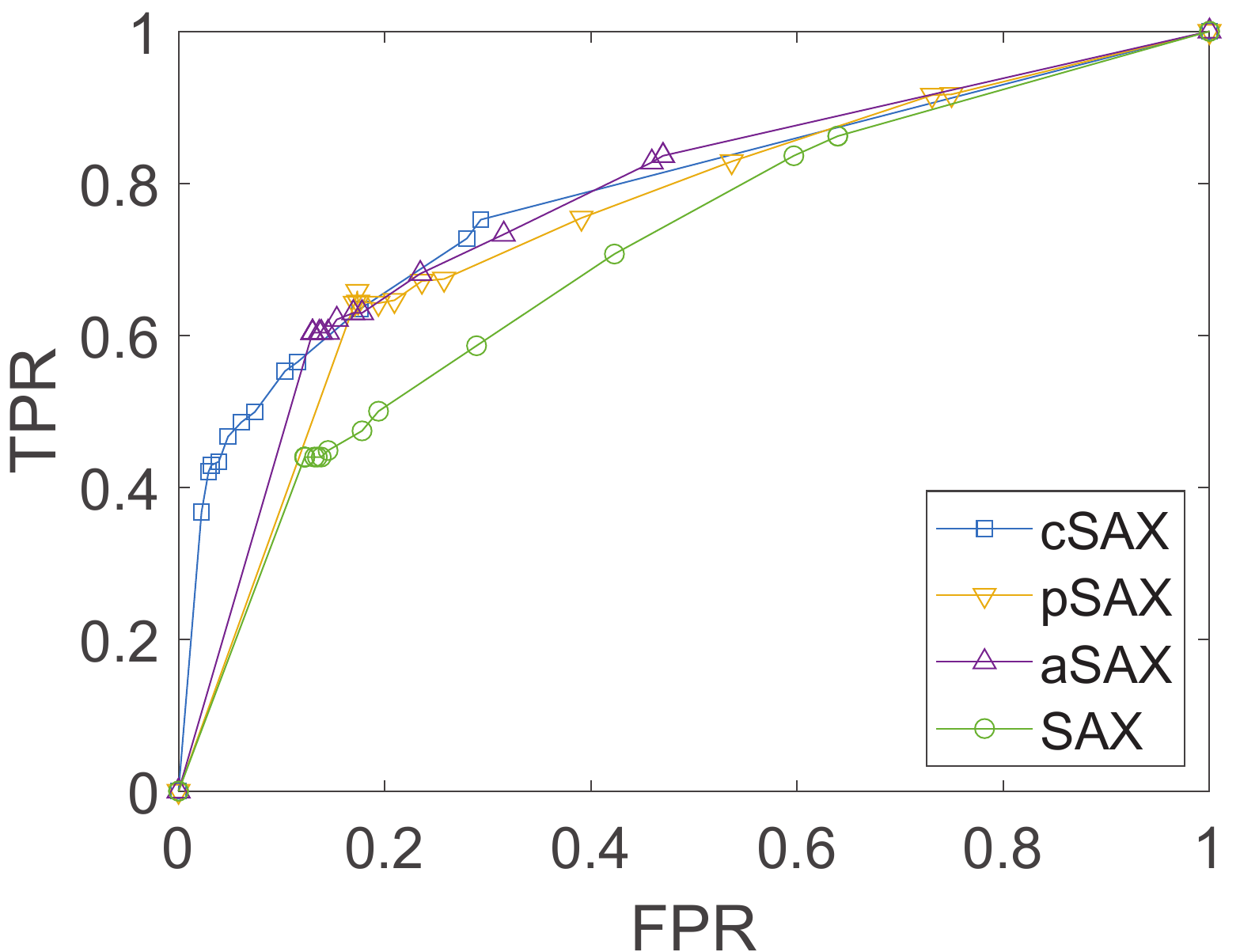}%
    }
    \hfil
    \subfloat[]{\includegraphics[width=.5\linewidth]{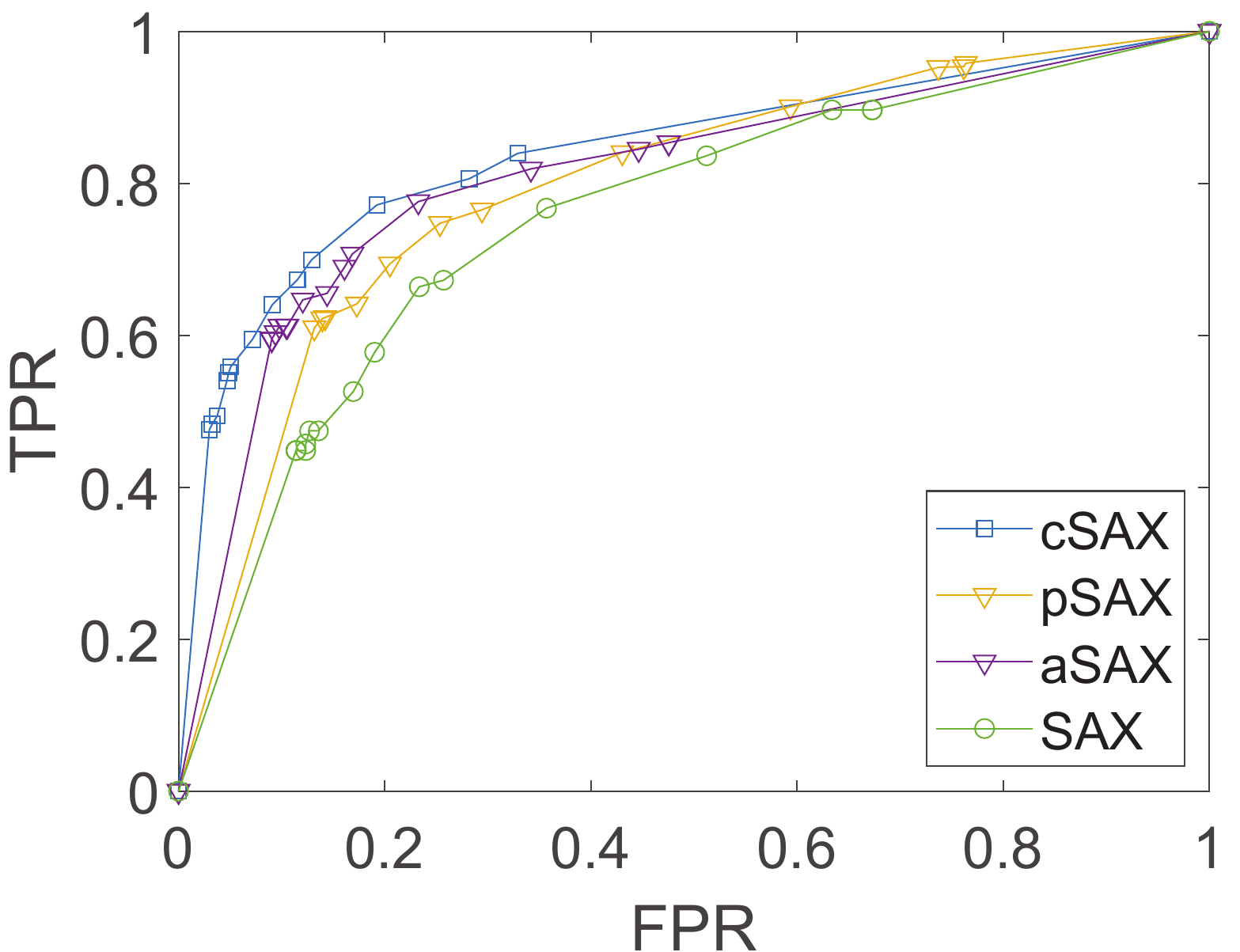}%
    }\\ \vspace{-7pt}
    \subfloat[]{\includegraphics[width=.5\linewidth]{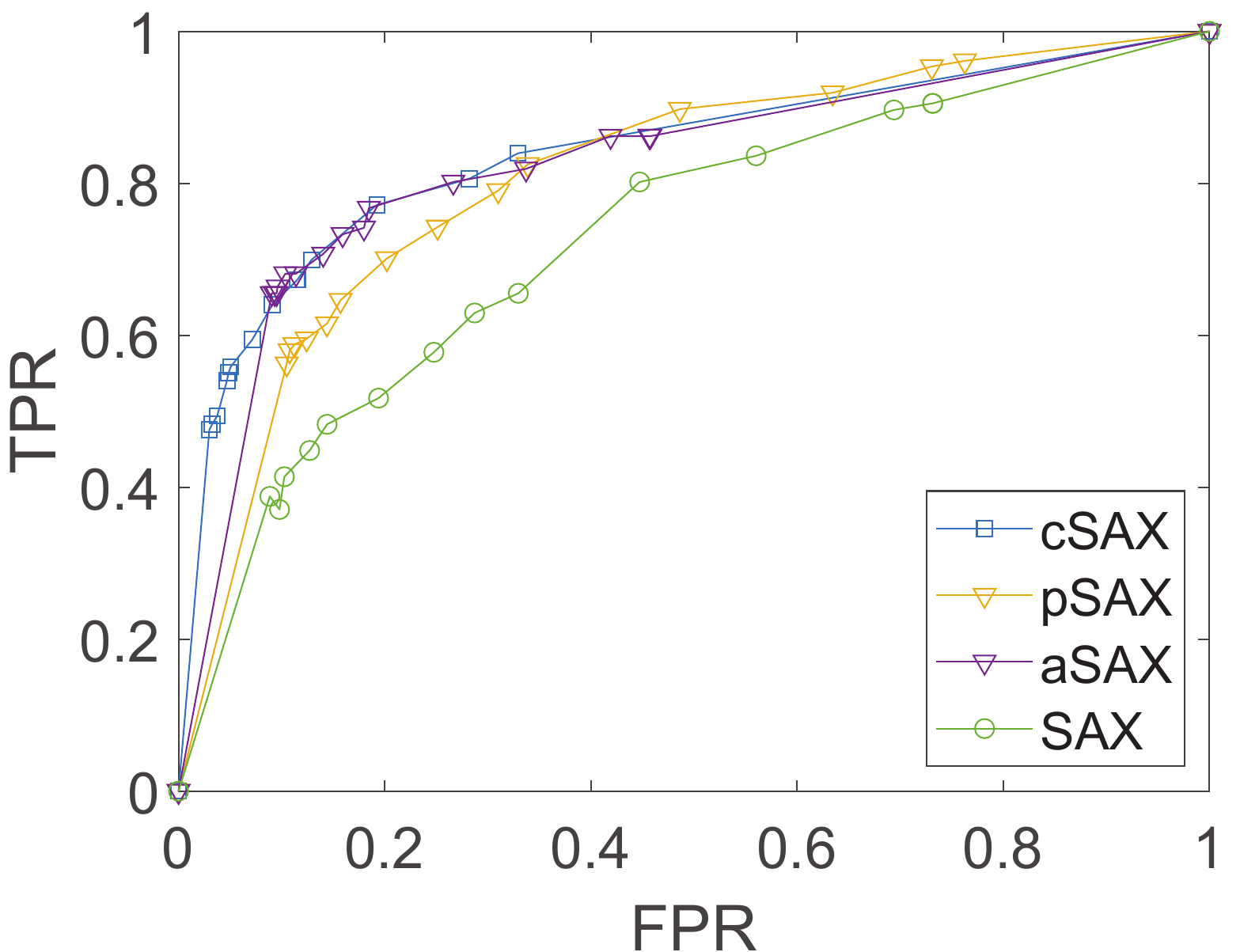}%
    }
    \hfil
    \subfloat[]{\includegraphics[width=.5\linewidth]{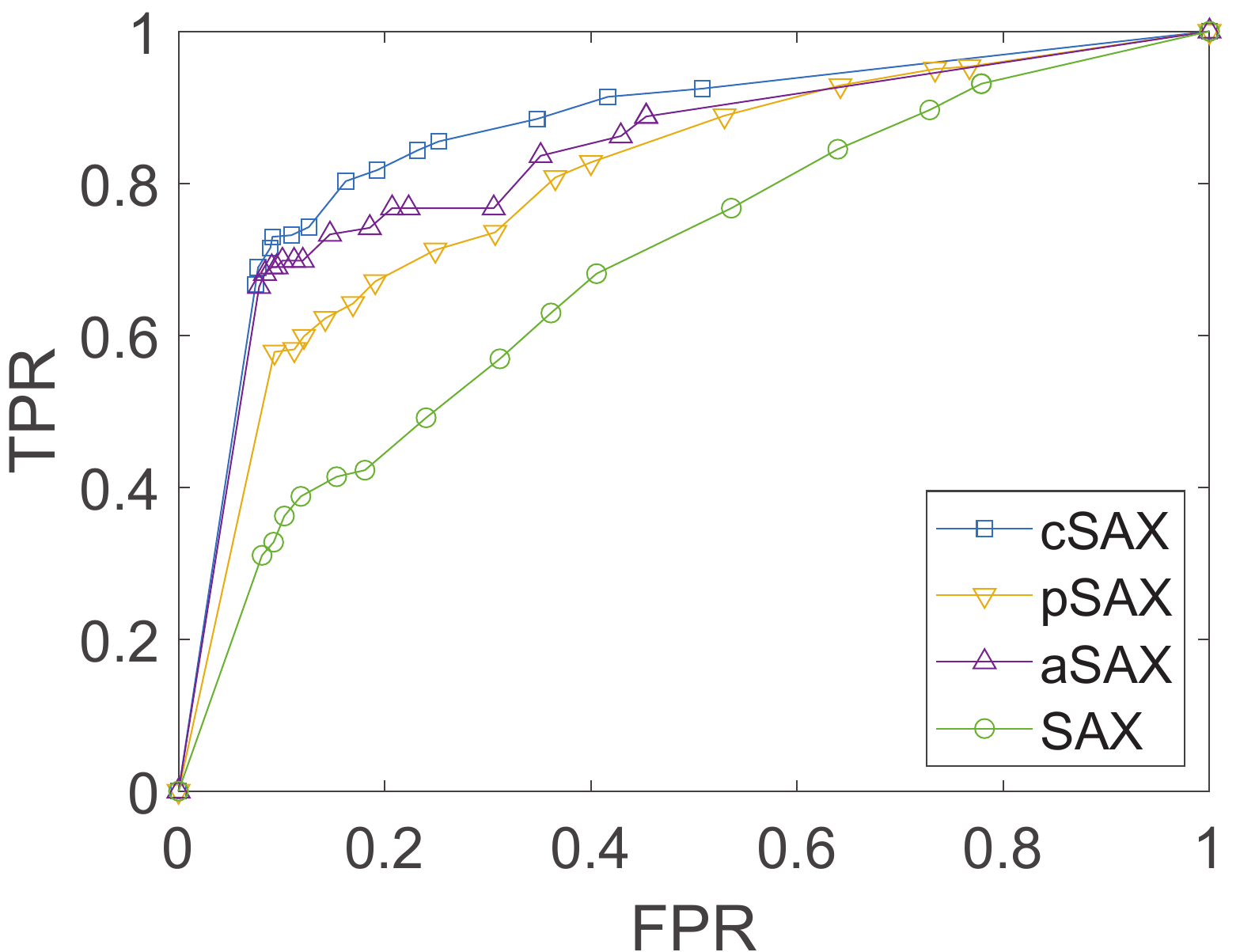}%
    }\\
    \smallskip\smallskip
    \begin{tabular}{|l |c|c|c|c|}
    \hline
    \footnotesize{M/N} & \footnotesize{cSAX} & \footnotesize{pSAX} & \footnotesize{aSAX} & \footnotesize{SAX}\\
    \hline
    \footnotesize{(a) 1/32} & \footnotesize{$\mathbf{0.7870}$} & \footnotesize{0.7551} & \footnotesize{0.7716} & \footnotesize{0.7053}\\
    \hline
    \footnotesize{(b) 1/16} & \footnotesize{$\mathbf{0.8413}$} & \footnotesize{0.7979} & \footnotesize{0.8113} & \footnotesize{0.7580}\\
    \hline
    \footnotesize{(c) 1/8} & \footnotesize{$\mathbf{0.8634}$} & \footnotesize{0.8132} & \footnotesize{0.8296} & \footnotesize{0.7358}\\
    \hline
    \footnotesize{(d) 1/4} & \footnotesize{$\mathbf{0.8703}$} & \footnotesize{0.8046} & \footnotesize{0.8394} & \footnotesize{0.6950}\\
    \hline
    \end{tabular}
    \caption{ROC curves (top) and AUC values (bottom table) of the SAX-based representations on NAB's datasets for 100\% training set size and different degrees of dimensionality reduction: a) $M/N = 1/32$, b) $M/N = 1/16$, c) $M/N = 1/8$, d) $M/N = 1/4$.}
    \label{fig:symbolic_ROC_with_dim}
\end{figure}

Fig.~\ref{fig:symbolic_ROC_with_dim} illustrates the effect of dimensionality reduction on the detector's performance. We hold the training set size to 100\% of the total samples and vary the degree of dimensionality reduction $M/N$ in $\{1/32, 1/16, 1/8, 1/4\}$ (from large to low degree of dimensionality reduction). As it can be seen, the conventional SAX has the worst performances, whereas cSAX clearly outperforms all the other SAX-based representations.

\subsubsection{Discord Discovery with cSAX}

Discord discovery with HOT-SAX is evaluated by the number of times the distance function is called, as explained in Sec.~\ref{sec:HOTSAX}. The datasets we employ belong to the most recent HexagonML/UCR Time Series Classification Archive\cite{bib:UCRAnomaly}, which includes 250 real-world labelled datasets. All of the datasets contain a single anomaly, hence they are appropriate for discord discovery algorithms. As in the previous subsection, some training sets are utilized, only for pre-training the SAX representations. For this evaluation, we employ directly the training sets already pre-defined in the archive. Since HOT-SAX works in a batch instead of an online fashion, the cSAX is trained on the pre-defined training set only, with no subsequent dynamic clustering.

\begin{figure}
    \centering
    \includegraphics[width=\linewidth]{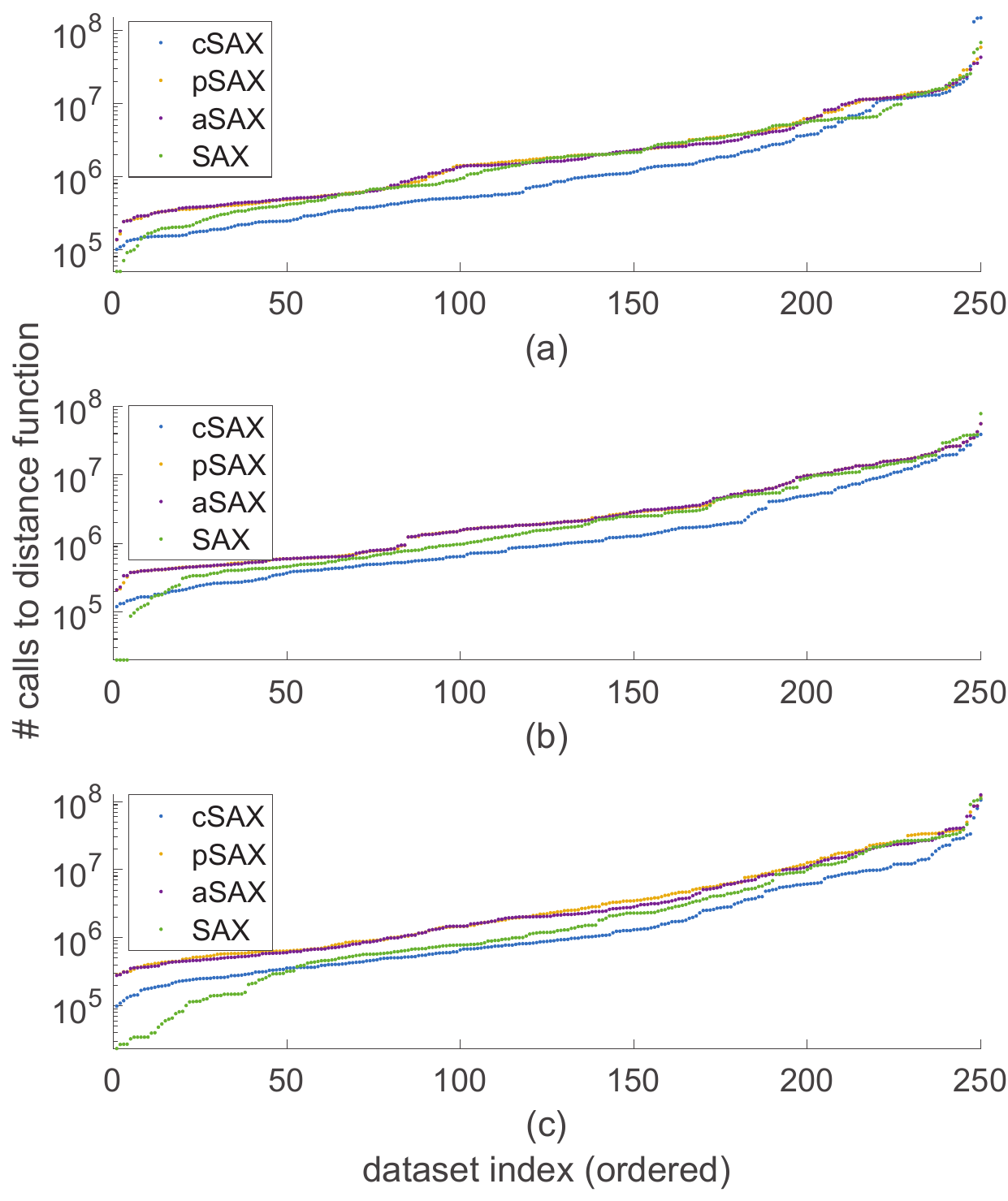}\vspace{-5pt}
    \caption{HOT-SAX running time in terms of number of calls to the distance function (the fewer the faster) for discord lengths: a) $l=64$, b) $l=128$, c) $l=256$.}
    \label{fig:HOTSAX_speed}
\end{figure}

Fig.~\ref{fig:HOTSAX_speed} illustrates the running time of HOT-SAX with each of the SAX representations for each of the datasets available in the aforementioned archive, when the discord length $l$ varies in $\{64, 128, 256\}$. To facilitate the visual comparison, the datasets are ordered in a scale from easier (faster) to harder (slower) processing.

We observe that for all discord lengths, the cSAX representation results in a faster discord discovery for the vast majority of the datasets, while the conventional SAX clearly outperforms pSAX and aSAX.

\section{Conclusions and Future Work}\label{sec:concl}
This paper proposed two novel non-parametric, data-driven symbolic representations, named pSAX and cSAX, by modifying the discretization process of the traditional SAX method. The proposed methods achieve to negate two restrictive assumptions of the traditional SAX, namely, the Gaussian distribution and the equiprobability of the symbols. The derived representations are based on kernel techniques to precisely adapt to the underlying data statistics. Furthermore, the pSAX representation employs the Lloyd-Max quantizer for general data compaction, whilst cSAX employs the mean-shift clustering for high-level time series analysis. A dynamic clustering mechanism was also introduced, such that the cSAX-based representation performs in a fully online fashion, enhancing its quality and simplifying its application for online data mining tasks. Equally importantly, both the pSAX and cSAX preserve the lower-bounding property and were shown to outperform the conventional SAX, as well as the aSAX, an alternative data-driven SAX-based representation. Notably, both pSAX and cSAX can be combined with the modifications of all other SAX-based representations, except for those that alter the discretization step, thereby increasing their performance.

In addition, we proposed two novel distance measures, which are not lower-bounding, but are optimal in the sense of mean squared error. Finally, we verified theoretically -- and proposed a solution thereupon -- a previously observed phenomenon that reduces the variance at the intermediate piecewise aggregate approximation, causing an additional information loss when left unnoticed.

The experimental evaluation of the methods was performed on a large collection of heterogeneous datasets. The experiments revealed that pSAX produces symbolic representations that are most often of higher quality than its competitors, whereas the superiority is more apparent in the settings of small alphabet size. The proposed cSAX was evaluated on the task of anomaly detection, both with and without dimensionality reduction. It was shown that the dynamic criterion is able to boost the performance of the anomaly detector, such that very good performance is achieved even when the clusters are not pre-trained. At the same time, both the non-dynamic and the dynamic mean-shift outperformed the alternative SAX-based representations for the purpose of anomaly detection. Very importantly, cSAX eliminates the need to tune the symbolic alphabet size, as it is automatically detected.

Currently, the pSAX method has not been tested in computationally demanding indexing scenarios. The paradigm of iSAX~\cite{bib:isax} shows that scalable indexing is possible when the discretization scheme is hierarchical. Motivated by this, we are interested in extending the Lloyd-Max quantizer in a hierarchical framework. Early experimentation shows that this is possible with virtually no decrease in the tightness of lower bound (TLB). At the same time, the proposed cSAX has not been evaluated in terms of TLB. It is of interest to see how the dynamic mean-shift affects the TLB, due to the varying quantization intervals. An extension of both pSAX and cSAX can be considered for the case of multi-channel time series, where the different channels are correlated. In that case, optimal discretization can be achieved by multivariate kernel density estimators for cSAX and vector quantizers for pSAX.


%

\appendix[Proof of Theorem \ref{thm:wrong_code_continuous}]\label{proof:thm:wrong_code_continuous}

\begin{proof}
Denote with $G(x_i)$ the pmf of the quantized $X$ under the wrong distribution $g(x)$. Assuming optimal coding under $G(x_i)$, the expected description length is given by (ref.~\cite[Thm. 5.4.1]{bib:Thomas})

\begin{equation}\label{eq:thm_proof}
\medmath{
\begin{aligned}
    \mathbb{E}[l(X^{\Delta})]    =& \sum_{x_i} P(x_i)\ceil{\log\frac{1}{G(x_i)}} \\
                        <& \sum_{x_i}P(x_i)\left(\log\frac{1}{G(x_i)}+1\right)\\
                        =& \sum_{x_i}P(x_i)\left(\log\frac{P(x_i)}{G(x_i)}\frac{1}{P(x_i)}+1\right)\\
                        =& \sum_{x_i}P(x_i)\log\frac{P(x_i)}{G(x_i)} + \sum_{x_i}P(x_i)\left(\log\frac{1}{P(x_i)}+1\right)\\
                        =& \sum_{x_i}f(x_i)\Delta\log\frac{f(x_i)\Delta}{g(x_i)\Delta} + \sum_{x_i}f(x_i)\Delta\left(\log\frac{1}{f(x_i)\Delta}+1\right)\\
                        =& \sum_{x_i}f(x_i)\Delta\log\frac{f(x_i)}{g(x_i)} - \sum_{x_i}f(x_i)\Delta\left(\log(f(x_i)\Delta)-1\right)\\
                        =& \sum_{x_i}f(x_i)\Delta\log\frac{f(x_i)}{g(x_i)} - \sum_{x_i}f(x_i)\Delta\log f(x_i)\\ -& \sum_{x_i}f(x_i)\Delta(\log\Delta-1) \rightarrow \, D(f\parallel g) + h(X) + \log\frac{1}{\Delta} + 1\ ,\\
                        & \mathrm{as}\ \Delta\,\rightarrow\, 0\ ,
\end{aligned}}
\end{equation}
where Riemann integrability is assumed in the last line of~\eqref{eq:thm_proof}.

Similarly, it can be shown that
\begin{equation}
    \mathbb{E}[l(X^{\Delta})] \geq D(f\parallel g) + h(X) + \log\frac{1}{\Delta}\ ,\quad \mathrm{as}\ \Delta\rightarrow 0\ .
\end{equation}
\end{proof}


%
%



\bibliographystyle{IEEEtran}
\bibliography{IEEEabrv,main}

\end{document}